\documentclass[a4paper,11pt]{article}
\usepackage{jcappub} 
\usepackage{amsmath}
\usepackage{bm}
\usepackage{caption}
\usepackage{lineno}
\usepackage{subcaption}
\usepackage{hyperref}
\usepackage{orcidlink}

\title{\boldmath Probing the end of reionization with 21cm--[CII] line intensity mapping
cross-correlations}

\author[a]{Shiriny Akthar \orcidlink{0009-0008-6537-2143},}
\author[b,c]{Tirthankar Roy Choudhury \orcidlink{0000-0001-7462-8587},}
\author[a]{Leon Noble \orcidlink{0009-0004-3138-1130},}
\author[a]{Suman Majumdar \orcidlink{0000-0001-5948-6920},}
\author[a]{Manas Mohit Dosibhatla \orcidlink{0000-0003-4229-2972},}
\author[a]{Yashrajsinh Mahida \orcidlink{0009-0000-1796-797X},}
\author[d]{Chandra Shekhar Murmu \orcidlink{0000-0002-1818-5440}}

\affiliation[a]{Department of Astronomy, Astrophysics and Space Engineering, Indian Institute of Technology Indore, Khandwa Road, Simrol, Indore 453552, India}
\affiliation[b]{National Centre for Radio Astrophysics, TIFR, Pune University Campus, Ganeshkhind, Pune 411007, India}
\affiliation[c]{International Centre for Theoretical Sciences, TIFR, Bengaluru 560089, India}
\affiliation[d]{Astrophysics Research Center of the Open University (ARCO) and Department of Natural Sciences, The Open University of Israel, 1 University Road, Ra’anana 4353701, Israel}

\emailAdd{shirinyakthar@gmail.com}
\abstract{The end of the epoch of reionization (EoR) marks a crucial transition in the thermal and ionization history of the intergalactic medium (IGM), yet its precise timing remains poorly constrained. We investigate the prospects of identifying the end of reionization through the cross-correlation between the redshifted neutral hydrogen (HI) 21cm signal from both the IGM and dark matter halos and the [CII] line emission tracing star-forming galaxies. We have developed a modified version of the \texttt{ReionYuga} simulation that incorporates the contribution of the 21cm signal coming from the galaxies hosted in dark matter halos through a halo-based HI prescription, along with the 21cm signal from the IGM, and generates 21cm and [CII] line intensity mapping (LIM) fields throughout the late stages of reionization. We characterize the evolving relationship between the two tracers using the Pearson correlation coefficient (PCC) and the 21cm--[CII] LIM cross-power spectrum, which further captures the scale-dependent correlation between the two fields. We find that the cross-correlation evolves from predominantly anti-correlated to correlated as ionized regions grow and eventually merge. This crossover of the 21cm–[CII] LIM cross-correlation arises from the competition between an anti-correlated diffuse-IGM component and a positively correlated halo-associated HI component, with the scale dependence of their crossover encoding information about the evolving morphology of the ionized IGM. We have shown that the transition to a positive cross-correlation on all scales considered provides a robust signature of the completion of reionization.
As a consistency check of our inferred end-of-reionization redshift, we estimate the mean free path (MFP) of ionizing photons in the late stages of reionization and assess whether the MFP of the ionizing photons is comparable to the length of the simulation box as the entire IGM is ionized. 
Additionally, we examine the detectability of the cross-power spectrum in upcoming 21cm observations by SKA-Low and [CII] LIM observations by FYST. 
This opens a new window to pinpoint the end of reionization.
}


\begin{document}
\newcommand{\actaa}{Acta Astron.} 
\newcommand{\araa}{Annu. Rev. Astron. Astrophys.} 
\newcommand{\aar}{Astron. Astrophys. Rev.} 
\newcommand{\ab}{Astrobiol.} 
\newcommand{\aj}{Astron. J.} 
\newcommand{\apj}{Astrophys. J.} 
\newcommand{\apjl}{Astrophys. J. Lett.} 
\newcommand{\apjs}{Astrophys. J. Suppl. Ser.} 
\newcommand{\ao}{Appl. Opt.} 
\newcommand{\apss}{Astrophys. Space Sci.} 
\newcommand{\aap}{Astron. Astrophys.} 
\newcommand{\aapr}{Astron. Astrophys. Rev.} 
\newcommand{\aaps}{Astron. Astrophys. Suppl.} 
\newcommand{\baas}{Bull. Am. Astron. Soc.} 
\newcommand{\caa}{Chinese Astron. Astrophys.} 
\newcommand{\cjaa}{Chinese J. Astron. Astrophys.} 
\newcommand{\cqg}{Class. Quantum Gravity} 
\newcommand{\gal}{Galaxies} 
\newcommand{\gca}{Geochim. Cosmochim. Acta} 
\newcommand{\icarus}{Icarus} 
\newcommand{\jcap}{J. Cosmol. Astropart. Phys.} 
\newcommand{\jgr}{J. Geophys. Res.} 
\newcommand{\jgrp}{J. Geophys. Res.: Planets} 
\newcommand{\jqsrt}{J. Quant. Spectrosc. Radiat. Transf.} 
\newcommand{\memsai}{Mem. Soc. Astron. Italiana} 
\newcommand{\mnras}{Mon. Not. R. Astron. Soc.} 
\newcommand{\nat}{Nature} 
\newcommand{\nastro}{Nat. Astron.} 
\newcommand{\ncomms}{Nat. Commun.} 
\newcommand{\nphys}{Nat. Phys.} 
\newcommand{\na}{New Astron.} 
\newcommand{\nar}{New Astron. Rev.} 
\newcommand{\physrep}{Phys. Rep.} 
\newcommand{\pra}{Phys. Rev. A} 
\newcommand{\prb}{Phys. Rev. B} 
\newcommand{\prc}{Phys. Rev. C} 
\newcommand{\prd}{Phys. Rev. D} 
\newcommand{\pre}{Phys. Rev. E} 
\newcommand{\prl}{Phys. Rev. Lett.} 
\newcommand{\psj}{Planet. Sci. J.} 
\newcommand{\planss}{Planet. Space Sci.} 
\newcommand{\pnas}{Proc. Natl Acad. Sci. USA} 
\newcommand{\procspie}{Proc. SPIE} 
\newcommand{\pasa}{Publ. Astron. Soc. Aust.} 
\newcommand{\pasj}{Publ. Astron. Soc. Jpn} 
\newcommand{\pasp}{Publ. Astron. Soc. Pac.} 
\newcommand{\rmxaa}{Rev. Mexicana Astron. Astrofis.} 
\newcommand{\sci}{Science} 
\newcommand{\sciadv}{Sci. Adv.} 
\newcommand{\solphys}{Sol. Phys.} 
\newcommand{\sovast}{Soviet Ast.} 
\newcommand{\ssr}{Space Sci. Rev.} 
\newcommand{\uni}{Universe} 

\maketitle
\flushbottom

\section{Introduction}
\label{Intro}
The epoch of reionization (EoR) marks the last major phase transition in the thermal and ionization history of the Universe, during which the intergalactic medium (IGM) evolved from a fully neutral state to a highly ionized state. This transformation is driven by the ionizing radiation from the first generation of luminous sources, whose formation is itself seeded by the growth of cosmological density perturbations \citep{Barkana_2001, Loeb_2001}. Reionization therefore occupies a unique position in cosmic history. It is driven by astrophysical processes operating in individual galaxies and halos, yet its effects are imprinted on the IGM across cosmological volumes. Therefore, it is a natural bridge between astrophysics and cosmology \citep{Morales_2010, Gnedin_2022}.

The EoR also connects two very different epochs of cosmic structure: the nearly homogeneous Universe seen in the cosmic microwave background (CMB) and the highly inhomogeneous large-scale structure (LSS) observed at low redshift \citep{Furlanetto_2006}. Characterizing this intermediate stage is essential not only for understanding the nature of the first ionizing sources, but also for understanding how their feedback shaped subsequent structure formation \citep{Barkana_2001, Choudhury_2006}.

Despite substantial observational and theoretical progress, several fundamental questions about reionization remain open, including the identity and relative contribution of the dominant ionizing sources, the timing and duration of reionization, the growth rate of ionized regions, and — the focus of this work — the redshift at which reionization completes \citep{Robertson_2015}. Pinning down the end of reionization is particularly valuable, as it provides important constraints on the ionizing photon budget of the source population and the late-time evolution of the IGM \citep{Robertson_2015, Becker_2015}.

A variety of observational probes currently constrain different stages of reionization: Ly-$\alpha$ absorption in quasar spectra, which traces the residual neutral hydrogen (HI) fraction along individual lines of sight \citep{Becker_2015, Becker_2001, Fan_2006, Barnett_2017, Bandos_2018, Gaikwad_2020, Choudhury_2021, Bosman_2022, Spina_2024}; the Thomson scattering optical depth of CMB photons, which constrains the redshift-integrated ionization history \citep{komatsu_2010, Planck_2016, Planck_2018}; the abundance and clustering of Ly-$\alpha$ emitters (LAEs), which are sensitive to the surrounding IGM opacity \citep{Iliev_2007, Hu_2010, Morales_2021, Tang_2023, Kakiichi_2023, Witten_2024}; and, more recently, the 21cm forest as a probe of reionization of the IGM \citep{Furlanetto_2_2006, Yidong_2009, Soltinsky_2021, Shimabukuro_2025, Kongprachaya_2026, Soltinsky_2026}. Collectively, these probes have established increasingly tight constraints on the timing and duration of reionization. However, each of these techniques relies on observing a limited number of sightlines or sources. As a result, they carry limited statistical power and cannot directly trace the three-dimensional distribution of ionized and neutral regions in the IGM \citep{Pritchard_2011, Furlanetto_2006, Morales_2010}.
The Ly-$\alpha$ forest exhibits spatial correlations exceeding 200 Mpc in its transmission spikes at z$\sim$6 \citep{Spina_2026}, but recent studies demonstrate that this correlation is disproportionately driven by a small number of highly transmissive sightlines \citep{Saptarshi_2026}. This motivates the need for a complementary technique that can directly map the spatial structure of the high-redshift Universe throughout the EoR.

Line intensity mapping (LIM) has emerged as a powerful technique for this purpose. Rather than resolving individual sources, LIM measures the aggregate emission from a given spectral line across large cosmological volumes, enabling efficient three-dimensional tomography of the high-redshift Universe \citep{Kovetz_2017, Kovetz_2_2019, Bernal_2022, chang_lidz_2026_lim}. Several emission lines have been proposed as tracers of this era, including the HI 21cm line \citep{Jonathan_2008, Pritchard_2011, TRC_2016}, the [CII] 158$\mu$m fine-structure line \citep{Gong_2012, Yue_2015, Lagache_2018, Yue_2019, Sun_2021, Bethermin_2022, Moradinezhad_2022, Yang_2022, Kannan_2022, Padmanabhan_2022, G.Sun_2023, Van_2023, Karoumpis_2022, Karoumpis_2024, Liang_2024, Liu_2024}, rotational transitions of carbon monoxide (CO) \citep{Carilli_2011, Lidz_2011, Mashian_2015, Breyesse_2017, Breysse_2022, Chung_2023, Dosibhatla_2026}, Ly-$\alpha$ $\&$ H$\alpha$ emission \citep{Silva_lyman_2013, Pullen_2014, Heneka_2017, Silva_2016, Visbal_2018, Chang_2019, Heneka_caroline_2021, Mas-Ribas_2023, Guochao_2023, Pad_2024} and many more lines. Several completed, ongoing, and planned ground-based, balloon-borne, and space-based experiments aim to deliver LIM measurements spanning a wide range of cosmic epochs, including the  Giant Metrewave Radio Telescope (GMRT) \citep{Govind_1991}, Precision Array to Probe the Epoch of Reionization (PAPER) \citep{Parsons_2010}, LOw
Frequency ARray (LOFAR) \citep{Haarlem_2013}, Murchison Widefield Array (MWA) \citep{Tingay_2013},  Hydrogen Epoch of Reionization Array (HERA) \citep{David_2017} and Square Kilometre Array Observatory (SKAO) \cite{Koopmans_2015} targeting the 21cm line, and Fred Young Submillimeter Telescope (FYST) \citep{CCAT_2023}, CarbON CII line in post-rEionisation and ReionisaTiOn epoch (CONCERTO) \citep{CONCERTO_2020}, Tomographic Ionized-carbon Mapping Experiment (TIME) \citep{TIME}, etc. targeting the [CII] line. 

The use of multiple line intensity tracers is particularly valuable because different emission lines respond to different physical environments. The 21cm line traces neutral IGM, and the [CII] line is sensitive to astrophysical processes within the interstellar medium (ISM), including star formation, gas cooling, chemical enrichment, and feedback. Cross-correlating LIM fields from distinct tracers can therefore provide information unavailable from either field considered independently \citep{Bernal_2022}. In the 21cm and [CII] fields, the two tracers are expected to exhibit a characteristic evolution in their spatial relationship as reionization proceeds. In the early stages, ionized regions form and expand around [CII] line-emitting galaxies, while substantial amounts of HI remain in the diffuse IGM. The corresponding 21cm and [CII] fluctuations are therefore expected to be spatially anti-correlated \citep{Fronenberg_2024}. As ionized regions grow and merge, the HI in the diffuse IGM is progressively depleted. Toward the end of reionization, the remaining HI becomes increasingly associated with galaxies and their surrounding environments \citep{Bharadwaj_2001}, which are also traced by [CII] emission. The cross-correlation between HI 21cm and [CII] can consequently evolve from negative to positive values, providing a characteristic indication of the changing relationship between the two tracers. This evolving relationship can be quantified using statistical measures of the two LIM fields. The cross-correlation measures the degree to which fluctuations in the 21cm and [CII] fields are spatially correlated \citep{Roy_2026, Gagnon_2025}. 

Previous studies have investigated the cross-correlation between 21cm and [CII] LIM fields for different aspects of reionization and galaxy formation \citep{Fronenberg_2024, Roy_2026}. Such studies demonstrate the complementary nature of the two tracers and the potential of their cross-correlation to connect the distribution of HI with the population of star-forming galaxies. In particular, the transition from negative to positive cross-correlation reflects a shift in the origin of the HI traced by the 21cm field, from predominantly diffuse HI in the IGM to HI increasingly associated with galaxies and their surrounding environments. A further advantage of the cross-power spectrum is that it retains the scale dependence of this relationship. A one-point real-space correlation statistic provides an overall measure of the correlation between the two fields, whereas the cross-power spectrum describes how this correlation varies with Fourier scale. The cross-power spectrum is also sensitive to the relative phase of the Fourier modes of the two fields. It therefore contains information about their relative spatial configuration at different scales. This is particularly relevant during reionization, when the characteristic sizes and spatial distribution of ionized regions evolve with time. The cross-power spectrum can therefore distinguish the scale-dependent evolution of the relationship between the HI distribution and the [CII] line-emitting galaxy population, providing information beyond what a single integrated correlation measure contains.

In this work, we exploit LIM to develop an alternative tomographic approach for identifying the end of reionization, based on the cross-correlation between two complementary tracers: the HI 21cm emission and the [CII] line emission. The 21cm signal traces HI in the diffuse IGM during the EoR, and transitions to tracing self-shielded HI within halos once reionization completes \citep{Pritchard_2011}. The [CII] line, in contrast, originates from the ISM of star-forming galaxies hosted by dark matter (DM) halos throughout this period \citep{DeLooze_2014, Herrera_2015}. These two tracers therefore probe different components of the evolving reionization process. 
Their combination consequently provides a way to relate the evolution of the neutral IGM to the spatial distribution of the galaxies embedded within it. We first characterize the overall relationship between the 21cm brightness temperature and [CII] LIM fields using the Pearson correlation coefficient (PCC), which provides a one-point statistical measure of their correlation. We then extend the analysis to the two-point statistic via the 21cm--[CII] cross-power spectrum, which allows the evolution of the correlation to be studied as a function of Fourier scale. We find that the correlation between the two fields evolves markedly during the late stages of reionization. The PCC changes from negative to positive values as the neutral IGM becomes progressively ionized,  while the cross-power spectrum reveals how this transition develops across different spatial scales. We use the scale‑dependent transition in the 21cm--[CII] cross‑correlation to define the end of reionization in our fiducial reionization and [CII] models. As an independent validation of the inferred end-of-reionization redshift, we compute the mean free path (MFP) of ionizing photons and investigate whether it approaches the scale of the simulation box toward the end of reionization.

An important aspect of this approach is its potential observational feasibility. Both 21cm and [CII] emission can be mapped tomographically, allowing their spatial correlation to be studied over a sequence of redshifts rather than through an integrated measurement. Moreover, because the two lines are observed in different frequency bands and with independent instrumental systems, their cross-correlation is less sensitive to contaminants that are uncorrelated between the two measurements. This provides a useful advantage over statistics based on a single intensity map, although correlated astrophysical and instrumental systematics must still be carefully accounted for in a realistic analysis. We therefore investigate whether the characteristic transition identified in our simulations can be detected in upcoming observations, such as those from SKA-Low and FYST. To assess the observational feasibility of this method, we construct instrumental noise realizations corresponding to survey configurations representative of the SKA-Low for the 21cm line and the FYST for [CII]. We propagate the corresponding thermal noise into the 21cm--[CII] cross-correlation and estimate the detectability of the cross-power spectrum.
This allows us to determine whether the characteristic evolution of the cross-power spectrum found in our simulations can be recovered with statistical significance via SKA-Low and FYST observations.

This paper is organized as follows. In section \ref{Simulation}, we describe the LIM simulations used to construct the 21cm brightness temperature and [CII] LIM fields. We also summarize the survey specifications considered for the detectability analysis. In section \ref{Methodology}, we describe the statistical methods used to characterize the relationship between the two fields, including the PCC and the cross-power spectrum, and discuss the procedure for estimating the signal's observational detectability. In section \ref{Results}, we present our
results. We first investigate the evolution of the 21cm--[CII] PCC and then study the auto- and cross-power spectra of the two fields. We subsequently present the detectability estimates for the 21cm--[CII] LIM cross-correlation across the survey configurations considered. Finally, we summarize our main findings and present our conclusions in section \ref{Summary}.

Throughout this work, we assume a flat $\Lambda$CDM cosmology with $h=0.6766$,  $\Omega_{\rm m}=0.2621$,  $\Omega_{\rm b}=0.0490$, and $\Omega_{\Lambda}=0.6888$ \citep{Planck_2_2018}. All distances are quoted in comoving units.

\section{Line intensity mapping simulation}
\label{Simulation}
\subsection{21cm brightness temperature maps}
We use a semi-numerical simulation based on the excursion-set formalism \citep{Furlanatto_2004} to generate the redshifted 21cm fields during the EoR. The simulation consists of the following steps:
\begin{itemize}
    \item We generate the DM field using a dark-matter–only particle-mesh (PM) \texttt{N-body}\footnote{\url{https://github.com/rajeshmondal18/N-body}} simulation \citep{Somnath_2004, Mondal_2015}. The DM field is simulated on a $3072^3$ grid with a spatial resolution of $0.07 ~\mathrm{Mpc}$, which corresponds to a simulation volume of [$215.04$\, \rm Mpc$]^3$ and a mass resolution of $1.08 \times 10^8 M_\odot$.
    \item In the next step, we identify DM halos using the \texttt{Friends-of-Friends}\footnote{\url{https://github.com/rajeshmondal18/FoF-Halo-finder.git}} algorithm, with a fixed linking length of $0.2$ times the mean inter-particle separation \citep{Mondal_2_2015}. We consider a halo to be resolved if it contains at least 10 DM particles, which corresponds to a minimum halo mass of $1.08 \times 10^9 M_\odot$.
    \item We apply an excursion-set based algorithm to the DM field and the halo catalog to generate the ionization field using the semi-numerical code \texttt{ReionYuga}\footnote{\url{https://github.com/rajeshmondal18/ReionYuga}} \citep{TRC_2009, Suman_2014, Mondal_3_2017}. It has three free parameters. The first parameter, $M_{(h, min)}$, sets the minimum halo mass that can host ionizing sources and participate in reionization. The second parameter, $N_{ion}$, quantifies the number of ionizing photons per baryon, assuming that the number of ionizing photons produced by a source is proportional to the host halo mass ($M_h$). The third parameter, $R_{mfp}$, represents the mean free path of ionizing photons in the IGM. This framework assumes that the hydrogen density traces the underlying DM density and DM halos host sources capable of producing ionizing photons. The values for $M_{h, min}$, $N_{ion}$ and $R_{mfp}$ are $1.08 \times 10^9 M_{\odot}$, $24.71$ and $17.24\, \rm Mpc$, respectively.
    
    \item To incorporate the contribution of HI from the DM halos to the total 21cm emission during the end stages of reionization, we modify the standard \texttt{ReionYuga} simulation. The \texttt{upgraded ReionYuga (uReionYuga)} includes the contribution of HI associated with halos in addition to the diffuse IGM component. In this modified implementation, we assign HI to DM halos using the HI mass ($M_{\rm HI}$) to halo mass ($M_h$) relation from  \citep{Navarro_2018},
    \begin{equation}
M_{\rm{HI}}(M_h) = M_{0} \left( \frac{M_h}{M_{\rm{min}}} \right)^{\alpha} \exp \left[ -(M_{\rm{min}}/M_h)^{0.35} \right].
\end{equation}
    We adopt the HI-halo mass relation calibrated at $z\sim5$, with parameter values $M_0=1.9\times10^{9}\,h^{-1}M_\odot$, $M_{\rm min}=2\times10^{10}\,h^{-1}M_\odot$ and $\alpha=0.74$. Here, $M_0$ fixes the normalization, corresponding to approximately $40\%$ of the HI mass associated with a halo of mass $M_{\rm min}$; $M_{\rm min}$ sets a low-mass cutoff for hosting HI and $\alpha$ determines the high-mass scaling.
\end{itemize}
On the $384^3$ coarse-grid with  $0.56 ~\rm Mpc$ resolution, both the IGM and halo HI densities are computed using Cloud-in-Cell (CIC) interpolation. The two are then summed in HI density, and the total HI density is converted into a 21cm brightness temperature map following equation (7) of \citep{Pritchard_2011};
\begin{equation}
\delta T_{\rm b}
=
27 x_{\rm HI}
\left(1+\delta_{\rm b}\right)
\left(1-\frac{T_{\rm CMB}}{T_{\rm S}}\right)
\left(\frac{\Omega_{\rm b}h^2}{0.023}\right)
\left(\frac{0.15}{\Omega_{\rm m}h^2}\frac{1+z}{10}\right)^{1/2}
\,\mathrm{mK}.
\label{21cm_brightness_temperature}
\end{equation}
Here $x_{\rm HI}$ is the HI neutral fraction, and $z$ specifies the redshift of the region emitting the 21cm signal. The $\delta_b$ and $T_{\rm CMB}$ denote baryonic density contrast and the CMB temperature at redshift $z$, respectively. This approach allows us to construct 21cm fields that consistently include contributions from both the residual HI in the IGM and the HI associated with DM halos, shown in figure \ref{field_map_2}, which is essential for reliably probing the end stage of reionization through cross-correlation analyses.

\subsection{[CII] line intensity maps}
\begin{figure}
    \centering
    \includegraphics[width=1\linewidth]{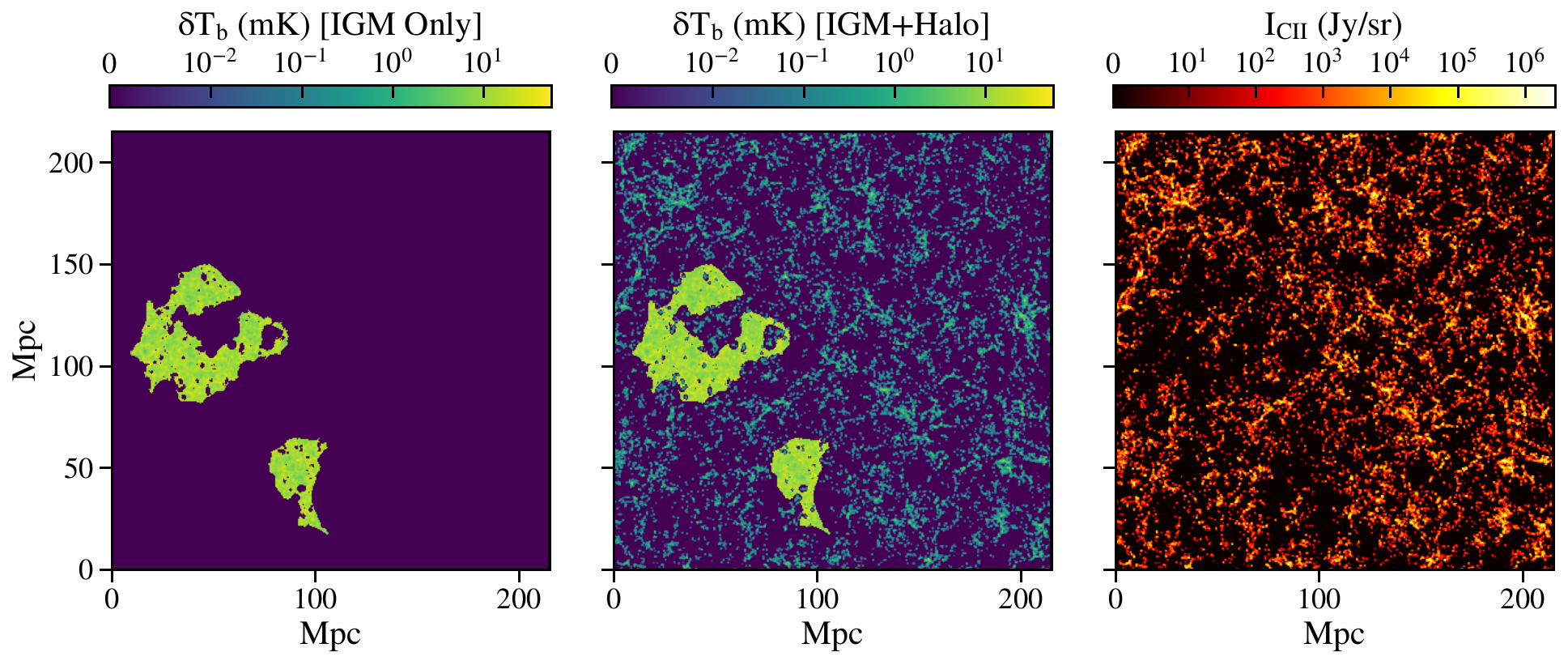}
   \caption{\textbf{Left:} 21cm brightness temperature map of the IGM. \textbf{Middle:} 21cm brightness temperature map including contributions from both the IGM and HI associated with dark matter halos. 
   \textbf{Right:} [CII] LIM map. Each subplot shows a $0.56$ Mpc thick slice of the corresponding map at $z\sim6$ on a logarithmic color scale.}
    \label{field_map_2}
\end{figure}

The [CII] LIM fields are generated using the \texttt{LIMiT}\footnote{\url{https://github.com/dmmohit/LIMiT}} \citep{Dosibhatla_2026}, which is an updated version of \texttt{LIM\_simulator}\footnote{\url{https://github.com/chandra-001/LIM_simulator}} \citep{Murmu_2021}. In this approach, [CII] luminosities are assigned to the DM halos identified in the N-body simulation \citep{Somnath_2004, Mondal_2015}. To perform this, we adopt a set of parametric scaling relations that connect the halo mass to the [CII] luminosity.
\begin{itemize}
    \item First, the halo mass is related to the star-formation rate (SFR) of the corresponding halo using;
   \begin{equation}
\begin{aligned}
\frac{\rm{SFR}}{M_{\odot} \rm{yr}^{-1}} = & 2.25 \times 10^{-26} \left(1 + (z - 7) 7.5 \times 10^{-2}\right) \\
& \times M^{a}_{h} \left(1 + \frac{M_{h}}{c_{1}}\right)^{b} \left(1 + \frac{M_{h}}{c_{2}}\right)^{d} \left(1 + \frac{M_{h}}{c_{3}}\right)^{e}
\label{sfr}
\end{aligned}
\end{equation}

with $a = 2.59$, $b = -0.62$, $d = 0.4$, $e = -2.25$, $c_{1} = 8 \times 10^{8} M_{\odot}$, $c_{2} = 7 \times 10^{9} M_{\odot}$, $c_{3} = 10^{11} M_{\odot}$ \citep{Silva_2013}.

    \item We then convert the SFR into the [CII] luminosity using the scaling relation:
    \begin{equation}
\log_{10} \left( \frac{L_{\rm{[CII]}}}{L_{\odot}} \right) = a_{L_{\rm{[CII]}}} \times \log_{10} \left( \frac{\rm{SFR}}{M_{\odot} \rm{yr}^{-1}} \right) + b_{L_{\rm{[CII]}}}
\label{lum}
\end{equation}
   For this work, we considered model $m_1$ ($a_{L_{\rm CII]}} = 0.8475$, $b_{L_{\rm [CII]}} = 7.2203$), as this gives the highest [CII] intensity among four models \citep{Silva_2015}.
    \item Using this prescription, we construct a [CII] luminosity distribution from the simulated DM halo catalog. This distribution is then mapped onto a coarse three-dimensional grid using CIC interpolation. Finally, the [CII] luminosity at each grid point is converted into [CII] intensity using equation (\ref{intensity_eq}) \citep{Silva_2015}, resulting in the [CII] LIM fields at each redshift, as shown in figure \ref{field_map_2}.
    \begin{equation}
        I_{\rm CII}=\frac{1}{\Delta V}\sum_i\frac{L_{\rm CII}(M_i,z)}{4\pi H(z)}\lambda_{\rm CII}
        \label{intensity_eq}
    \end{equation}
\end{itemize}

\begin{figure}
    \centering
    \includegraphics[width=0.5\linewidth]{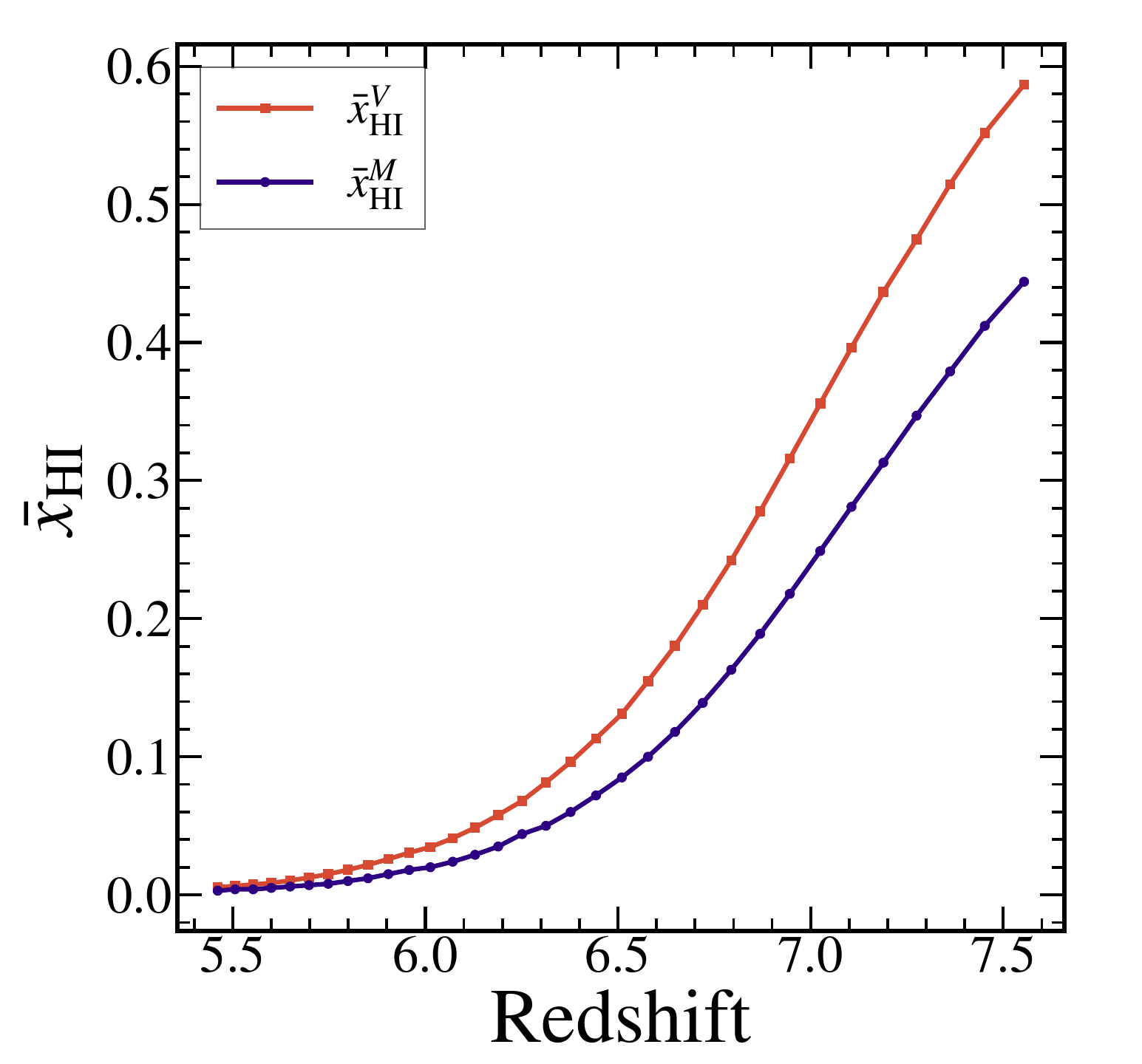}
    \caption{Neutral hydrogen fraction over the redshift range $5.5 \leq z \leq 7.5$. The red and blue curves show the volume-averaged ($\bar x^V_{HI}$) and mass-averaged ($\bar x^M_{HI}$) HI fractions, respectively.}
    \label{reionization_history}
\end{figure}

The simulation resolution is $\rm 0.56$ Mpc. We consider our analysis over the redshift range $z\sim [5.461 - 7.544]$, as our main goal is to identify the end of reionization. In figure \ref{reionization_history}, we show the reionization history over this redshift range.

\subsection{Surveys to detect 21cm and [CII] lines}
The Square Kilometre Array Low-Frequency instrument (SKA-Low) \citep{dewdney_2022}, under construction in Western Australia, will observe the redshifted 21cm emission from HI. SKA-Low will operate in the 50–350 MHz frequency range, corresponding to a redshift range of approximately $z \sim 3{-}28$ when interpreted as the observation of the HI 21cm signal, making it well suited for studies of the EoR and Cosmic Dawn (CD) \citep{Koopmans_2015}.
The Fred Young Submillimeter Telescope (FYST) \citep{CCAT_2023}, located in northern Chile, will perform LIM by targeting [CII] emission from the ISM of galaxies. It operates over a wavelength range of 200–3,100 $\mu$m (220-410 GHz), corresponding to a redshift range of approximately $z \sim 3$–$8$ for the [CII] line, making it well suited for probing the galaxy population during the EoR \citep{Parshley_2018}.
Figure 1 of \citep{Sarkar_2026} shows the overlapping redshift range and $k$-modes of SKA and FYST. This overlapping region is of particular interest for our work, as it defines the redshift and scale at which the HI 21cm and [CII] cross-correlation measurement can be performed, and serves as the focus of our analysis.

\section{Methods}
\label{Methodology}
\subsection{Cross-correlation statistics}
\subsubsection{Pearson 
correlation coefficient}
\label{Pearson Correlation}
The PCC \citep{Yule_1911} is a statistical measure that quantifies the degree of linear correlation between two variables. In our analysis, we use the PCC to quantify the correlation between the 21cm brightness temperature and [CII] LIM fields. It measures the degree to which fluctuations in two fields are linearly related, defined as the covariance between the two fields normalized by the product of their standard deviations. We use the PCC as an initial estimator of the correlation between the 21cm and [CII] LIM fields, defined as;
\begin{equation} 
    \rm PCC =\frac{\sum\limits_{i}\left( \delta T_{b,i} - \bar{\delta T}_{b}\right)\left( I_{\nu,i} - \bar{I}_{\nu} \right)}{\sqrt{\sum\limits_{i}\left(\delta T_{b,i} - \bar{\delta T}_{b} \right)^{2}\;\sum\limits_{i}\left( I_{\nu,i} - \bar{I}_{\nu}\right)^{2}}}
    \label{eq:cross-var} 
\end{equation}
where the sums extend over all voxels in the corresponding LIM maps. Here, $\delta T_{b, i}$ denotes the 21cm brightness temperature in the $i$-th voxel, while $I_{\nu, i}$ denotes the [CII] specific intensity in the same voxel. The quantities $\bar{\delta T}_{b}$ and $\bar{I}_{\nu}$ represent the corresponding volume-averaged mean values. The denominator provides normalization by the standard deviations of the two fields, such that $\rm PCC$ ranges from $\rm -1$ to $\rm 1$. A value of $\rm PCC=-1$ ($\rm PCC=1$) corresponds to perfect anti-correlation (correlation), while $\rm PCC=0$ indicates the absence of a linear correlation. 

\subsubsection{ Auto- and cross-power spectrum}
\label{Cross Power}

The power spectrum is defined through the fluctuations of two fields as;
\begin{equation}
    \left\langle \delta \tilde{I}_a(k)\,\delta \tilde{I}_b^{\ast}(k') \right\rangle = V\,\delta_{k,k'}\,P_{a \times b}(k) 
\end{equation}
where $\delta \tilde{I}_a(k)$ and $\delta \tilde{I}_b(k)$ denote the Fourier transforms of the fluctuations in fields $A$ and $B$, respectively, $V$ is the volume over which the fields are defined, and $P_{a\times b}(k)$ is their cross- power spectrum. Replacing $b$ with $a$ yields the auto-power spectrum of the field $A$, and vice versa. The asterisk denotes complex conjugation, and $\delta_{\bm{k},\bm{k}'}$ is the Kronecker delta.

The cross-power spectrum is a statistical measure used to quantify the correlation between fluctuations in two different fields. In contrast to the auto-power spectrum, which characterizes the statistical properties of fluctuations within a single field, the cross-power spectrum describes how fluctuations in one field relate to those in another field as a function of spatial scale. It is particularly useful when two fields trace related underlying structures, as their cross-power spectrum can provide information about the degree and nature of their correlation.

The cross-power spectrum therefore describes the correlation between the two fields at a given Fourier mode, or equivalently, at a given spatial scale characterized by $k$. 
In general, the $P_{a \times b}(k)$ defined above is not necessarily a real quantity. However, its imaginary part cancels when the cross-power is spherically averaged over all modes in a shell of $|k|$, as A and B are real fields. The spherically averaged cross-power spectrum is therefore real, though it can be negative (positive) whether the fields are anti-correlated (correlated). 

The cross-power spectrum also retains information about the relative phase between the two fields, as it captures the complementary evolution of the two fields, which will essentially provide the signature of the transition from IGM-dominated 21cm signal to halo-dominated 21cm signal. If the Fourier space fluctuations are written as;
\begin{equation}
   \delta\tilde{I}_a(\bm{k}) = a(\bm{k})\exp\left[i\phi_a(\bm{k})\right],
\qquad
   \delta\tilde{I}_b(\bm{k}) = b(\bm{k})\exp\left[i\phi_b(\bm{k})\right],
\end{equation}
Then their cross-correlation can be written as;
\begin{equation}
    \left\langle\delta\tilde{I}_a(\bm{k})\delta\tilde{I}_b^{\ast}(\bm{k})\right\rangle = \left\langle a(\bm{k})b(\bm{k})\exp\left\{-i\left[\phi_b(\bm{k})-\phi_a(\bm{k})\right]\right\}\right\rangle.
\end{equation}
Here, $\phi_a(\bm{k})$ and $\phi_b(\bm{k})$ represent the Fourier-space phases of the two fields.

Thus, the cross-power spectrum, defined as the ensemble average of the product of the Fourier modes of the two fields, contains information about the relative phase between their fluctuations, in addition to their amplitudes.
To quantify the correlation strength or the relative phase difference between the LIM signals, we define the cross‑correlation coefficient as;
\begin{equation}
   r_{a\times b}(k) = \frac{P_{a\times b}(k)}{\sqrt{P_a(k)P_b(k)}},
\label{eq:cross_correlation}
\end{equation}
Here, $P_a(k)$ and $P_b(k)$ are the auto-power spectra of fields $A$ and $B$, respectively. The coefficient $r_{a\times b}(k)$ quantifies the degree of correlation between the two fields as a function of scale. Values of $r_{a\times b}(k)$ close to $1$ indicate positively correlated fluctuations, while values close to $-1$ indicate anti-correlated fluctuations. A value close to zero indicates a weak or negligible linear correlation at that scale.

In this work, we apply this formalism to the 21cm brightness temperature and [CII] LIM fields. The corresponding auto- and cross- power spectrum are, $P_{21\rm cm}$ $P_{\rm [CII]}(k)$, quantifies their scale-dependent correlation. For convenience, we work with the normalized auto- and cross- power spectrum;

\begin{equation}
    \Delta^2_{21\rm cm}(k) = \frac{k^3}{2\pi^2}\,P_{21\rm cm}(k),
\end{equation}

\begin{equation}
    \Delta^2_{\rm [CII]}(k) = \frac{k^3}{2\pi^2}\,P_{\rm [CII]}(k),
\end{equation}

\begin{equation}
    \Delta^2_{21\rm cm\times{\rm [CII]}}(k) = \frac{k^3}{2\pi^2}\,P_{21\rm cm\times{\rm [CII]}}(k).
\end{equation}

In contrast, $r_{21\rm cm\times{\rm [CII]}}(k)$ provides a normalized measure of the strength and sign of this correlation. The sample variance of the corresponding auto- and cross- power spectrum will be;
\begin{equation}
    \sigma^2_{\rm sample}(k) = \frac{P^2(k)}{N_{\rm modes}}
    \label{sample}
\end{equation}
Here, $P(k)$ is the relevant auto- or cross- power spectrum.

The use of cross-power spectra to characterize the statistical relationship between complementary cosmological tracers has been explored in several previous studies \citep{Isabella_2017, Schaan_2021, Roy_cross_2024, Sarah_cross_2025, Debanjan_cross_2026}.

\subsection{Mean free path estimation}
\label{Mean Free Path Estimation}

The MFP of ionizing photons provides a useful measure of the characteristic distance that photons can travel through ionized regions before encountering HI. We estimate the MFP using the ray-tracing method introduced by \citep{Mesinger_2007}, which employs a Monte Carlo approach to characterize the sizes and spatial distribution of ionized regions during reionization. The \texttt{ReionYuga} simulation provides a continuous ionization fraction field, $x_{\rm HII}$, with values ranging from $0$ to $1$. To apply the MFP method, this continuous field must be converted into a binary ionization field by adopting a threshold value of $x_{\rm HII}$. Following the commonly used threshold, we classify a cell as ionized if $x_{\rm HII}\geq 0.5$ and neutral otherwise \citep{sambit_2018}.

In this method, a starting position is selected randomly from the ionized regions of the simulation volume, and a propagation direction is chosen isotropically. A ray is then traced from this position until it reaches a neutral cell, at which point the distance traveled is recorded as the MFP. We apply this procedure to the binary ionization field, in which each cell is classified as either ionized or neutral. The procedure is repeated for $\rm 10^7$ randomly selected starting positions and propagation directions, producing a distribution of ray lengths that characterizes the MFP of ionizing photons within the ionized regions. The resulting distribution can then be used to quantify the evolution of the MFP with redshift. This approach provides a statistical measure of the propagation distances permitted by the topology of the ionization field. As reionization progresses, ionized regions expand and merge, allowing rays to travel greater distances before encountering neutral cells. The resulting evolution of the MFP therefore provides complementary information about the growth and connectivity of ionized regions during the EoR and also an independent validation for our inferred end-of-reionization redshift from cross-power spectrum analysis, as it will indicate whether the mean free path of the photons is comparable to the simulation box size at the end of reionization.

\subsection{Detectability estimation}
\label{Detectability Estimation}
To assess the observational feasibility of the 21cm--[CII] LIM cross-correlation signal, we model the thermal noise expected for SKA-Low observation of the redshifted 21cm line.
We adopt the SKA-Low telescope in its \texttt{AA$^\ast$} configuration, generated with \texttt{ska\_ost\_array\_config}\footnote{\url{https://gitlab.com/ska-telescope/ost/ska-ost-array-config}} \citep{SKA_ost_array_config}, pointed at $\mathrm{RA}=0^{\circ}$, $\mathrm{Dec}=-30^{\circ}$ \citep{Bonaldi_2025} . To minimize ionospheric phase error \citep{Chege_2022, Gan_2023} and primary-beam effects \citep{Chokshi_2024}, the observation is centered on the field's meridian transit, with the UTC start time computed from the local sidereal time at the SKA-Low site. From the simulated $(u,v)$ coverage of this pointing, we extract the baseline-length distribution $|\bm{u}|=|\bm{b}|/\lambda$ and bin it into annuli of width $\delta u$ (the shortest baseline) to obtain the baseline density $n(u)$, i.e.\ the number of baseline pairs per unit $uv$-area per unit observing time, normalized by the number of independent time samples set by the integration time and total observation time per night. The frequency-dependent system sensitivity, $A_{\rm eff}/T_{\rm sys}$, is obtained by interpolating the SKA-Low performance parameters from the SKA Observatory science performance specifications \citep{Braun_2019} at the observed frequency corresponding to the redshift of the 21cm signal.

The thermal noise power spectrum of a radio interferometer observing the redshifted 21cm line is then given by \citep{Bull_2015, CSM_2024}; 
\begin{equation}
P_{\rm N, \rm 21cm}(k, z) = \frac{\chi(z)^2\, r_{\nu}(z)\, \lambda_{21cm}(z)^4}
{\left(A_{\rm eff}/T_{\rm sys}\right)^2\, t_{\rm obs}\, n_{\rm pol}\, n(u)\, \nu_{21cm}} ,
\label{noise_power}
\end{equation}
where $\chi(z)$ is the comoving distance, $r_\nu(z)=c(1+z)^2/H(z)$ converts a frequency interval to a comoving line-of-sight distance, $\lambda_{21}(z)=21\,{\rm cm}\,(1+z)$ is the redshifted wavelength, $n_{\rm pol}=2$ is the number of polarizations, $t_{\rm obs}$ is the total observing time, and $k_\perp = 2\pi u/\chi(z)$.

The line-of-sight modes are obtained by Fourier transforming the frequency axis. For a frequency bandwidth $B$, the Fourier transform along the frequency direction yields delay modes $\eta$, which are related to the comoving line-of-sight wavenumber by;
\begin{equation}
   k_\parallel(z) = \frac{2\pi\,\nu_{21}\,H(z)}{c\,(1+z)^2}\,\eta .
\end{equation}
We form the full three-dimensional noise power spectrum on a $(k_\perp,k_\parallel)$ grid with $k=\sqrt{k_\perp^2+k_\parallel^2}$, and spherically average it into logarithmically spaced $k$-bins to obtain the one-dimensional thermal noise power spectrum $P_{\rm N, \rm 21}(k,z)$. 
Using this thermal noise power spectrum, we generate $50$ independent Gaussian noise realizations.

For generating the [CII] noise cube, we first compute the white-noise power for the FYST experiment by linearly interpolating between the white-noise power values tabulated at four redshifts in table 1 of \citep{CCAT_2023}. As FYST is a single-dish instrument, its thermal noise power spectrum is white (scale-invariant) in $k$, in contrast to the scale-dependent 21cm thermal noise described above. Using the interpolated noise power at the redshift of interest, $P_{N,\rm [CII]}(z)$, we generate $50$ independent Gaussian noise realizations.

By cross-correlating all possible pairs of the 21cm and [CII] LIM signal-plus-noise realizations, we obtain a total of $2500$ independent estimates of the cross-power spectrum. We have $50$ signal-plus-noise realizations for both the 21cm and [CII] signals. These realizations are used to estimate the statistical uncertainty, $\sigma_{\rm noise}$, associated with the measurement. Using the resulting variance, we compute the signal-to-noise ratio (SNR) for the 21cm auto-power spectrum, the [CII] auto-power spectrum, and the 21cm--[CII] cross-power spectrum as;
\begin{equation}
{\rm SNR}(k) = \frac{P(k)}{\sqrt{\sigma_{\rm sample}^2(k) + \sigma_{\rm noise}^2(k)}} ,
\label{snr}
\end{equation}
where $P(k)$ denotes the relevant estimates of auto- or cross-power spectrum from observations, $\sigma_{\rm noise}(k)$ is the noise-realization scatter described above, and $\sigma_{\rm sample}(k)$ is the sample variance arising from the finite number of Fourier modes averaged in each $k$-bin (\ref{sample}).

We note that SKA-Low and FYST do not overlap across all $k_\parallel$ and $k_\perp$ modes. While doing the SNR estimation, we are not taking this into account. This will underestimate the error due to the sample variance \citep{Agarwal_2026, Fronenberg_2024}. 

\section{Results}
\label{Results}
\subsection{Pearson correlation coefficient}
\label{Pearson}
\begin{figure}
    \centering
    \includegraphics[width=0.7\linewidth]{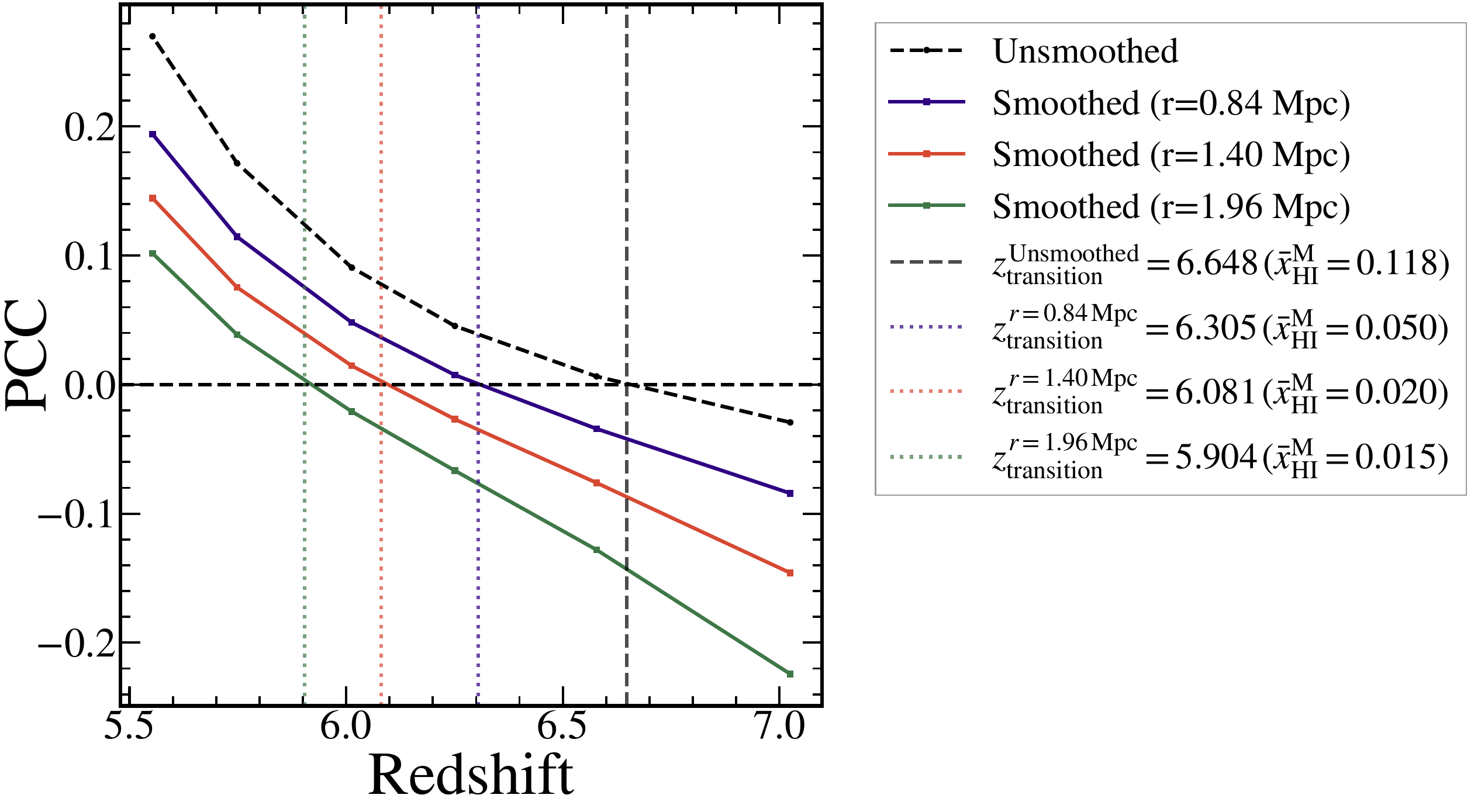}

    \caption{PCC as a function of redshift for different smoothing scales. The black dashed line shows the PCC at the original grid resolution of the simulation ($0.56$ Mpc). The indigo, red, and green solid curves correspond to smoothing scales of $0.84$, $1.40$, and $1.96$ Mpc, respectively. The corresponding dotted lines indicate the redshifts at which PCC transitions from negative to positive for each smoothing scale.}
    \label{cross_variance}
\end{figure}

The evolution of the correlation between the 21cm and [CII] fields is first studied using the PCC discussed in section \ref{Pearson Correlation}. 
Previous studies, e.g., \citep{Sarah_2026}, employed the PCC to probe the onset of reionization. In this work, we have employed PCC to probe the end of the EoR.
Figure \ref{cross_variance} shows the evolution of the PCC for different smoothing scales. The black dashed curve corresponds to the original simulation resolution of $0.56\,\rm{Mpc}$, while the indigo, red, and green solid curves correspond to spherical smoothing radii of $0.84$, $1.40$, and $1.96\,\rm{Mpc}$, respectively. The dotted vertical lines mark the redshifts at which PCC changes from negative to positive for the corresponding smoothing scales.

As star-forming galaxies emit photons that ionize the HI in the IGM, ionized regions form and grow around them. These ionized regions appear as cold spots in 21cm brightness temperature fields. Since the dominant source of the 21cm signal is from the IGM as compared to the HI associated with the galaxies, when reionization has not significantly progressed, the 21cm and CII LIM fields are anti-correlated, resulting in a negative PCC \citep{Adam_2009, Murmu_2021, Dumitru_2019}. As reionization proceeds, the neutral fraction of the IGM decreases, reducing the contribution of diffuse HI to the 21cm field. Consequently, the magnitude of the negative correlation decreases, and the PCC approaches zero toward the end of reionization. 
At later stages, the 21cm emission increasingly receives contributions from self-shielded HI associated with galaxies and their immediate environments, along with a simultaneous reduction in contribution to the same emission from HI in the IGM. Since these regions spatially overlap with
the [CII] line-emitting galaxy population, the correlation between the two fields changes sign and becomes positive \citep{Giri_2024}. This transition, therefore, serves as a characteristic indicator of the late stages of reionization and can be used to identify the end of reionization. 

We find that the transition in the sign of PCC depends on the smoothing scale. At the original resolution of $0.56\,\rm{Mpc}$, the transition occurs when the neutral fraction is approximately $10\%$. Increasing the smoothing radius shifts the transition toward progressively lower neutral fractions. For the largest smoothing scale considered, $1.96\,\mathrm{Mpc}$, the transition occurs at a neutral fraction of approximately $1\%$. This is happening because as we increase the smoothing scale, the residual 21cm signal in the IGM is smeared out over larger regions of the map, which destroys the positive correlation. Therefore, the crossover occurs later for larger smoothing scales. This scale dependence indicates that the transition in PCC does not correspond to a unique neutral fraction independent of spatial scale, but instead depends on the spatial scales over which the 21cm and [CII] fields are compared.

The PCC, therefore, provides a useful measure of the changing relationship between the two fields, while also revealing a clear dependence on the smoothing scale. To investigate this scale dependence more explicitly, we next consider the cross-power spectrum, which retains the correlation's dependence on Fourier scale and provides a complementary probe of the evolution of the 21cm--[CII] LIM cross-correlation toward the end of reionization.

\subsection{Two-point Fourier statistics}
\subsubsection{Auto-power spectrum}
\begin{figure}
    \centering
    \includegraphics[width=1\linewidth]{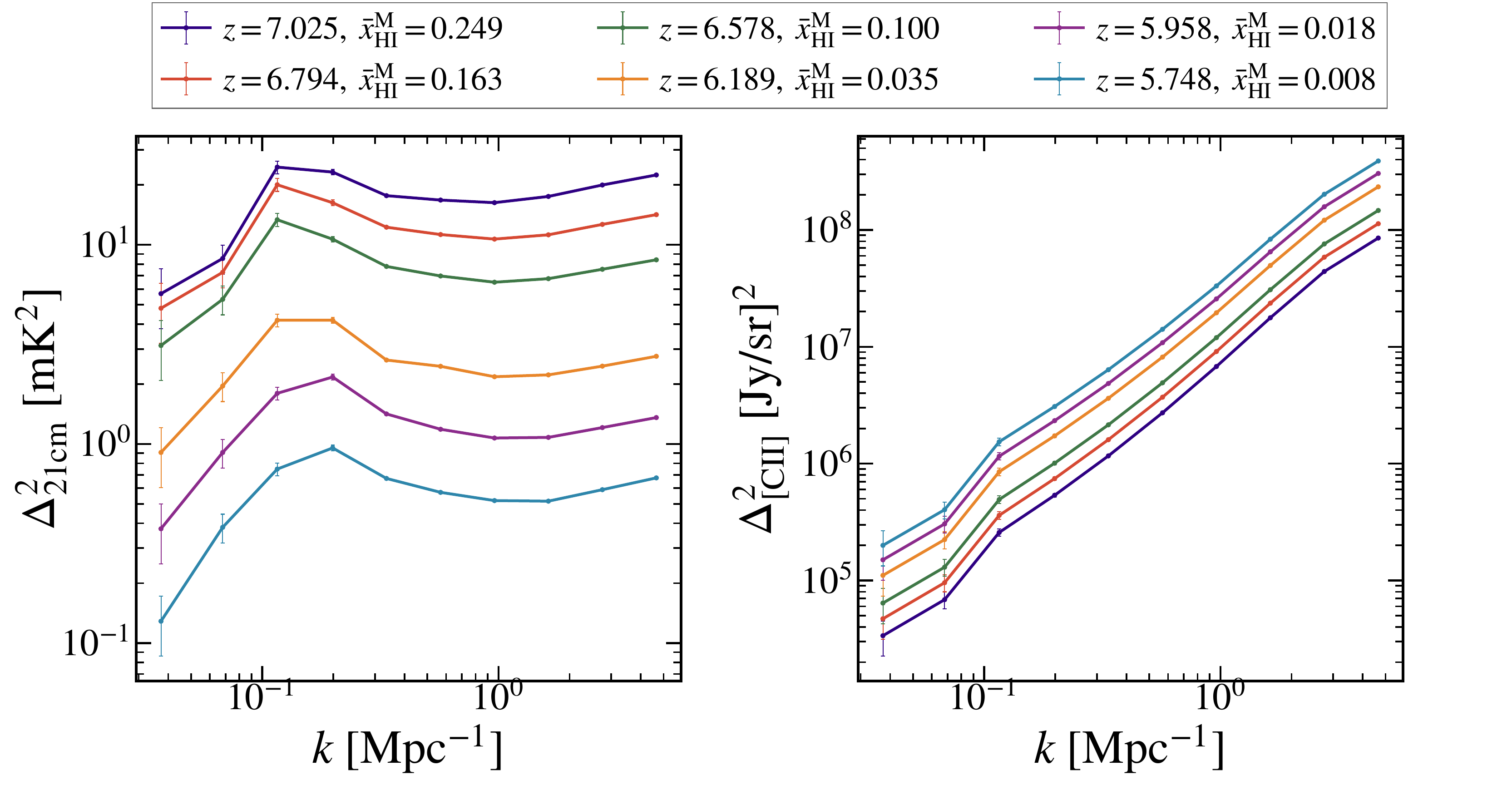}
    \caption{\textbf{Left:} 21cm power spectra as a function of wavenumber $k$, including contributions from both the IGM and halos, at redshifts $z=7.025\,, 6.794\,, 6.578\,, 6.189\,, 5.958$ and $5.748$, ordered from top to bottom. The corresponding mass-averaged HI fractions, $\bar x_{\rm HI}^M$, are indicated in the legend. \textbf{Right:} [CII] power spectra as a function of wavenumber $k$ at the same redshifts, ordered from bottom to top as the redshift decreases from $z=7.025$ to $z=5.748$. For both plots, the error bars denote sample variance.}
    \label{21cm_cii_ps}
\end{figure}

The 21cm brightness temperature power spectrum evolves significantly throughout reionization as the distribution of HI changes. The resulting evolution of the 21cm power spectrum over the redshift range considered in this work is shown in figure \ref{21cm_cii_ps}(Left).
At the beginning of reionization, the 21cm fluctuations are primarily determined by the spatial distribution of HI. As the first ionized regions emerge, the neutral fraction becomes inhomogeneous, modifying the 21cm fluctuations and leading to a suppression of its power spectrum on the relevant scales \citep{Lidz_2007, Georgiev_2021}. As reionization progresses, the ionized regions grow and merge. This enhances the 21cm brightness temperature fluctuations and causes the power spectrum to increase, eventually reaching a maximum. During this stage, the 21cm power spectrum is no longer determined solely by the underlying matter distribution, but is strongly influenced by fluctuations in the neutral fraction field. 
With further progression of reionization, the HI becomes increasingly confined to isolated regions as the ionized regions percolate and occupy an increasing fraction of the IGM. Consequently, the amplitude of the 21cm power spectrum decreases toward the end of reionization. A detailed discussion of the evolution and interpretation of the 21cm power spectrum can be found in \citep{Suman_2015}. 

We also compute the auto-power spectrum of the [CII] LIM field, shown in figure \ref{21cm_cii_ps}(Right). Since [CII] emission primarily originates from the ISM of star-forming galaxies, its spatial fluctuations trace the distribution and clustering of these galaxies. On large scales, the [CII] LIM field can therefore be described as a biased tracer of the underlying matter density field \citep{Dumitru_2019, Yue_2019, Bernal_2022}. In contrast to the 21cm power spectrum, the [CII] power spectrum exhibits a systematic increase in amplitude toward lower redshifts. This evolution reflects the growth of large-scale structure and the increasing clustering of galaxies as the Universe evolves. In addition, the increasing abundance of [CII] line-emitting galaxies and the evolution of their [CII] line luminosity with cosmic time enhance the amplitude of the intensity fluctuations \citep{Yue_2019, Karoumpis_2022}. Thus, the increase in the [CII] power spectrum toward lower redshifts results from the combined evolution of the underlying matter fluctuations and the population of
[CII] line-emitting galaxies. Also, the [CII] auto-power spectrum exhibits power-law behavior over the range of wavenumbers considered, with more power at higher $k$ because of stronger fluctuations on smaller spatial scales \citep{Dumitru_2019, Yue_2019, Karoumpis_2022}.
\subsubsection{Cross-power spectrum}
\label{Cross Power Results}
\begin{figure}
    \centering
    \includegraphics[width=1\linewidth]{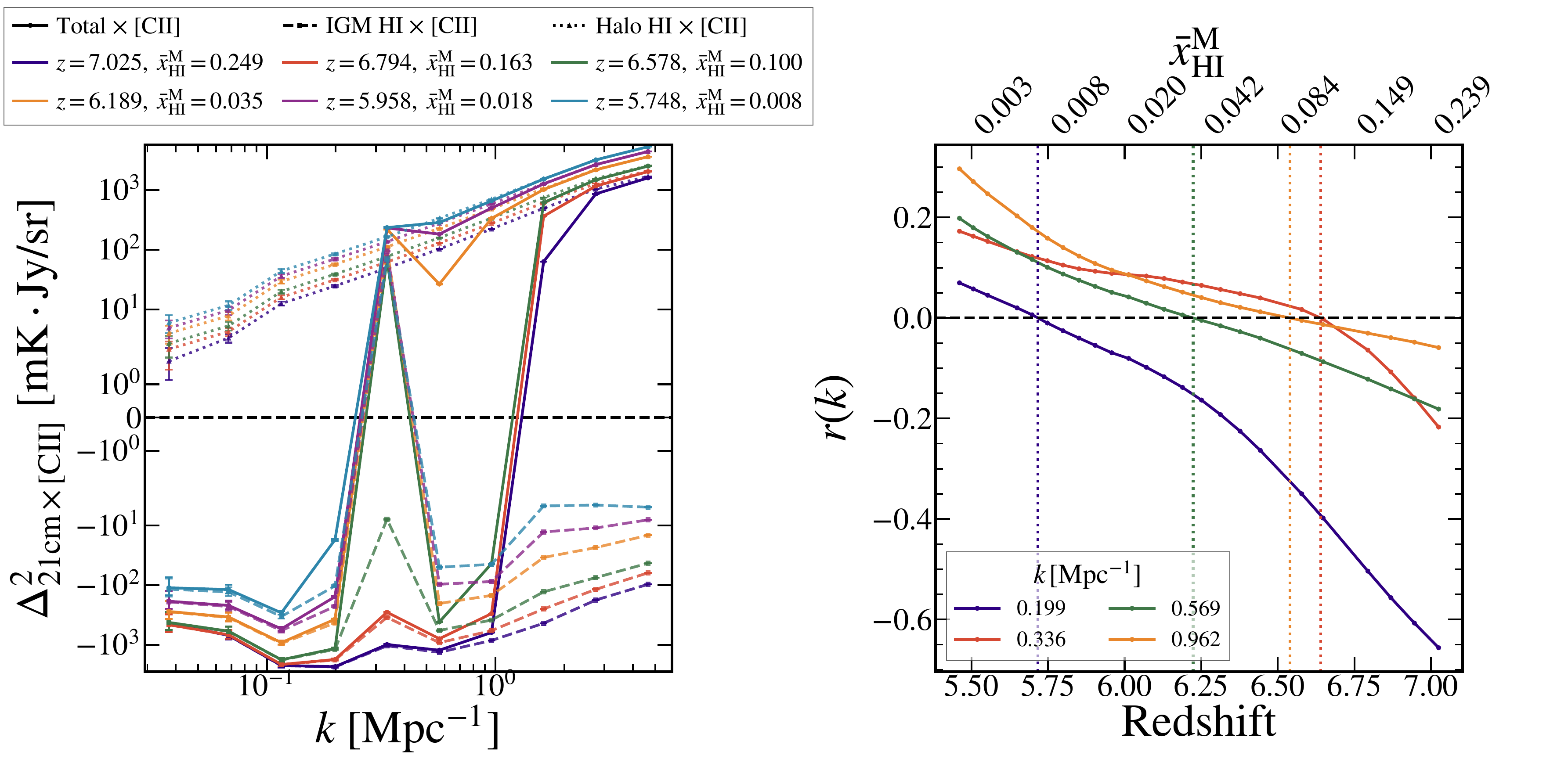}
    \caption{\textbf{Left:} 21cm--[CII] cross-power spectra as a function of wavenumber $k$ at redshifts $z=7.025$, $6.794$, $6.578$, $6.189$, $5.958$, and $5.748$, shown in blue, red, green, purple, orange, and teal, respectively. The error bars shown here denote sample variance. \textbf{Right:} Cross-correlation coefficient, $r(k)$, as a function of redshift for different wavenumber $k$.}
    \label{cross}
\end{figure}

The main aim of this work is to probe the end of reionization through the redshift evolution of the 21cm--[CII] cross-power spectrum.
Figure \ref{cross}(Left) shows the cross-power spectrum of the [CII] intensity and the 21cm differential brightness temperature as a function of Fourier mode, at several redshifts, obtained from our simulations. We consider the 21cm signal, including contributions from neutral hydrogen in both the diffuse IGM and DM halos, $\delta T_{b}=\delta T_{b}^{\rm IGM}+\delta T_{b}^{\rm halo}$, which we refer to as the total 21cm signal, and separately the contributions $\delta T_{b}^{\rm IGM}$ and $\delta T_{b}^{\rm halo}$. The corresponding normalized cross-correlation coefficient between [CII] and the total 21cm signal, $r(k)$, is shown in figure \ref{cross}(Right). Unlike the PCC, which provides a single measure of the overall correlation between the two fields, the cross-power spectrum retains information about their correlation as a function of Fourier mode. In particular, as discussed in section \ref{Cross Power}, the cross-power spectrum depends on the phase difference between the corresponding Fourier modes of the 21cm and [CII] LIM fields.
It therefore provides information about the relative spatial configuration of the two fields at different scales. This scale-dependent information is particularly relevant to the 21cm--[CII] LIM cross-correlation, since the spatial distributions of diffuse HI and [CII] line-emitting galaxies evolve differently across physical scales. The corresponding normalized cross-correlation coefficient further quantifies the degree of correlation at each Fourier scale after accounting for the individual auto-power amplitudes.
\begin{figure}
    \centering
    \includegraphics[width=1\linewidth]{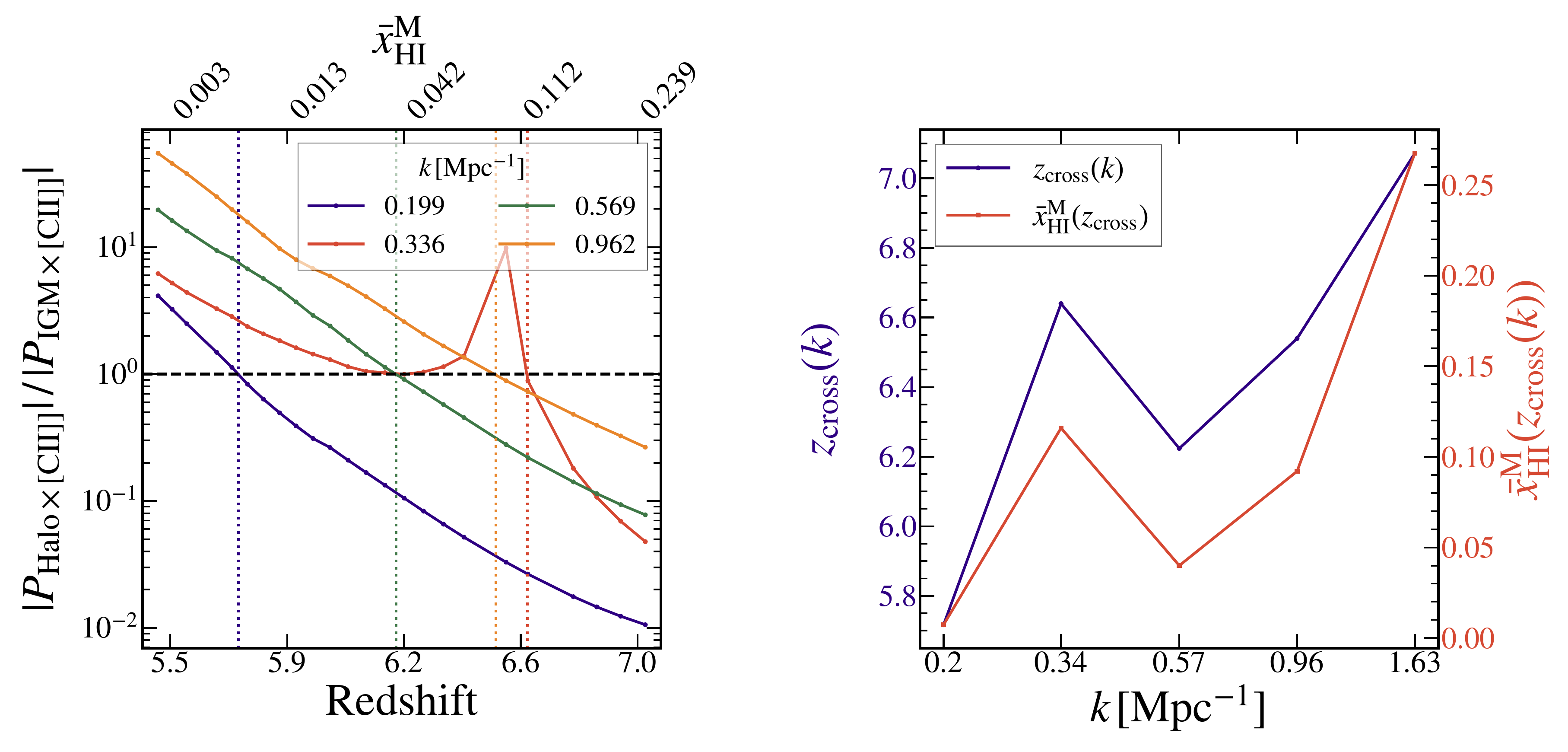}
    \caption{\textbf{Left:} The ratio $|P_{\rm halo\times[CII]}|/|P_{\rm IGM\times[CII]}|$ as a function of redshift for the $k$-modes that are detectable via future LIM experiments. \textbf{Right:} The scale-dependence of the crossover redshift $z_{\rm cross}(k)$, defined as the redshift at which the 21cm--[CII] cross-power spectrum changes sign from negative to positive at a given wavenumber $k$ (blue circles, left axis). The corresponding mass-averaged HI fraction $\bar{x}_{\rm HI}$ at $z_{\rm cross}(k)$ is shown for comparison (red squares, right axis), illustrating the stage of reionization associated with the sign transition at each scale.}
    \label{cross_2}
\end{figure}
During the early and intermediate stages of reionization, the 21cm field is dominated by HI in the diffuse IGM, whereas the [CII] LIM field traces [CII] line-emitting, star-forming galaxies. Since galaxies ionize the surrounding IGM, the progressively ionized regions around the galaxies are spatially anti-correlated with [CII] LIM fields. Consequently, the 21cm--[CII] cross-power spectrum has a negative amplitude over the range of $k$-modes considered during these stages of reionization. As reionization progresses, the ionized regions grow and merge, gradually reducing the contribution of diffuse HI to the 21cm field. The magnitude of the negative cross-power therefore decreases, and the cross-power spectrum approaches zero. At later stages, the cross-power changes sign from negative to positive at larger $k$-modes, reflecting the increasing contribution of HI associated with galaxies and their immediate environments. Once the diffuse IGM is almost completely ionized, the remaining HI is predominantly associated with galaxies. Since these regions overlap with the [CII] line-emitting galaxy population, the two fields become positively correlated, resulting in a positive cross-power spectrum. 

Figure \ref{cross_2}(Left) shows the redshift evolution of the ratio between $P_{\mathrm{halo}\times \mathrm{[CII]}}$ and $\left|P_{\mathrm{IGM}\times \mathrm{[CII]}}\right|$ for different $k$-modes. This tells us the redshift when the 21cm signal from halos will start dominating the 21cm signal from the IGM.
The first term ($P_{\mathrm{halo}\times \mathrm{[CII]}}$) is positive because both halo HI and [CII] are assigned to the same underlying halo population. The second term ($\left|P_{\mathrm{IGM}\times \mathrm{[CII]}}\right|$) should be predominantly negative during reionization, because galaxies preferentially reside in ionized regions and hence anti-correlate with diffuse HI.
The vertical dotted line represents the redshift $z_{\rm cross}(k)$ at which $P_{\mathrm{halo}\times \mathrm{[CII]}}(k) \simeq \left|P_{\mathrm{IGM}\times \mathrm{[CII]}}(k)\right|$.
The corresponding relation between $z_{\rm cross}$ at different k-modes is shown in figure \ref{cross_2}(Right). The largest (smallest) k-mode probes the smallest (largest) spatial scale, and therefore the $z_{\rm cross}$ at larger $k$ occurs at higher redshift.

The IGM 21cm--[CII] cross-power spectrum shows a localized positive feature at $k\simeq0.33\,\mathrm{Mpc^{-1}}$. Its physical origin is not understood. We have verified that it is not an artifact of the binning, of the particular realization, or of a few outlying modes; the diagnostic tests are described in Appendix \ref{Appendix}.

We define the end-of-reionization redshift, $z_{\rm end}$, as the redshift at which the 21cm--[CII] cross-power spectrum first changes sign at the smallest wavenumber considered, $k\simeq0.2\,\mathrm{Mpc}^{-1}$. We adopt this scale because it corresponds to the largest spatial scale accessible in our simulation volume for which the cross-power is expected to be detectable with future LIM experiments. At smaller wavenumbers, the measurements are expected to become sample variance dominated, limiting the detectability of the cross-power. We use this transition as our operational criterion to identify the redshift at which the IGM has reached the globally ionized state, according to our cross-power-based definition. With this criterion, we find $z_{\rm end}\simeq5.748$ for our reionization and [CII] models. At this redshift, the mass-averaged HI fraction of the IGM is $x_{\rm HI}^{\rm M}\simeq0.008$. This indicates that reionization has reached the end stage, with the remaining HI increasingly associated with galaxies and their surrounding environments. The positive cross-power at this scale reflects the positive spatial correlation between the 21cm and [CII] fields on large scales. We stress that $z_{\rm end}$ is defined through the cross-power spectrum and is specific to our adopted models, so it should not be interpreted as a model-independent measurement of the end of reionization. Our inferred end-of-reionization redshift aligns with existing constraints, e.g., Ly-$\alpha$ forest, LAEs, 21cm forest, and the Thomson scattering optical depth of CMB photons, with the reionization process completing between $z\sim 5.5{-}6$ \citep{Becker_2001, Fan_2006, Mortlock_2011, Schroeder_2013, McGreer_2015, Greig_2017, Spina_2026}.



\subsection{Mean free path estimates}
\label{MFP Results}
\begin{figure}
    \centering
    \includegraphics[width=1\linewidth]{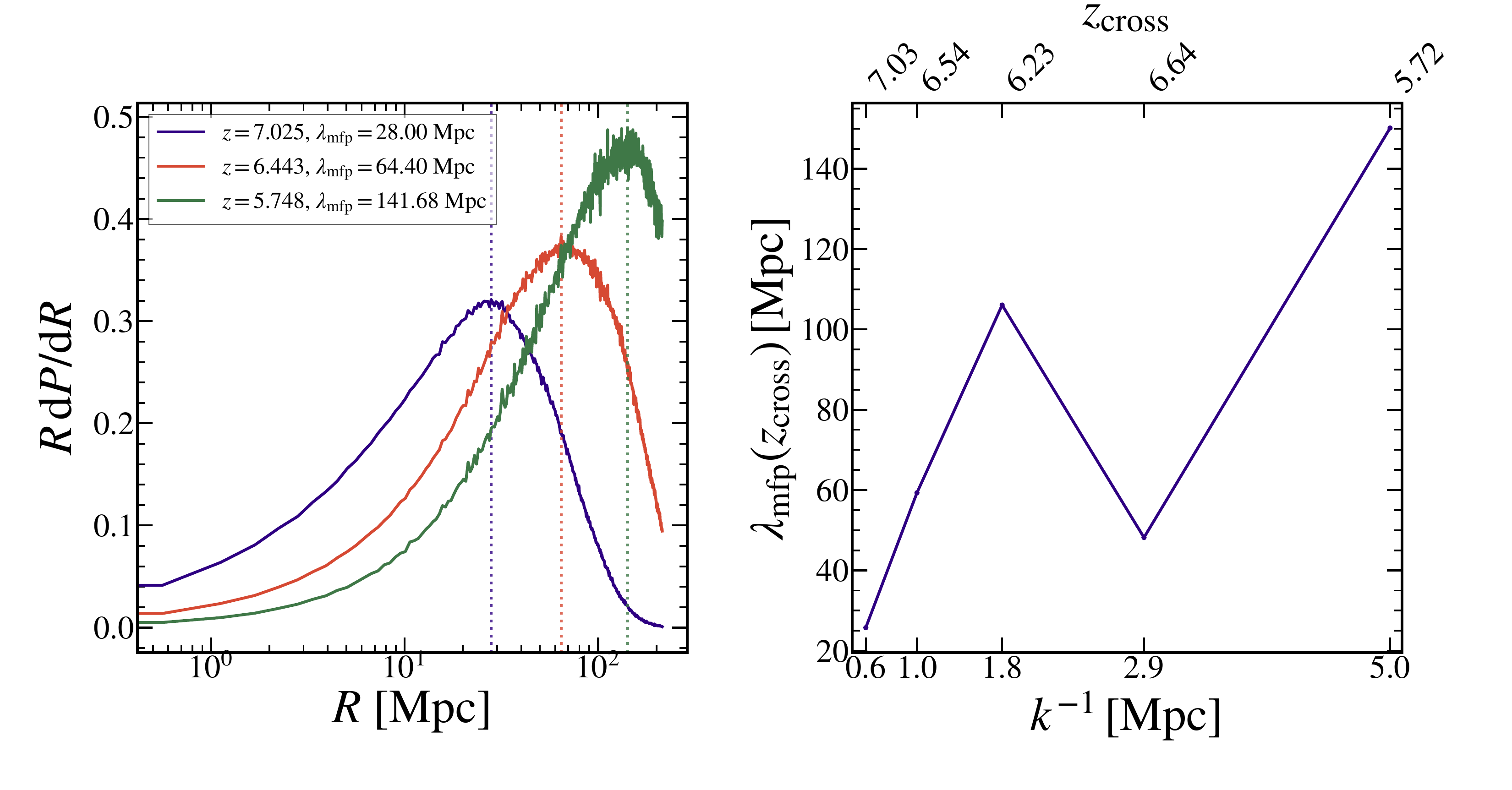}
     \caption{\textbf{Left:} Bubble size distributions (BSDs) at three different redshifts. The vertical lines indicate the peak positions of the distributions, with the corresponding bubble radii $R$ used to estimate the characteristic mean free path (MFP) of ionizing photons. \textbf{Right:} The mean free path of ionizing photons, $\lambda_{\rm mfp}$, evaluated at the transition redshift $z_{\rm cross}(k)$, as a function of the inverse wavenumber $k^{-1}$. Here $z_{\rm cross}(k)$ is the redshift at which the 21\,cm$\times$[CII] cross-power spectrum changes sign from negative to positive at a given $k$, and $\lambda_{\rm mfp}$ is taken as the peak of the bubble size distribution at that redshift. The corresponding value of $z_{\rm cross}$ for each scale is indicated on the top axis. This figure links each scale to the characteristic size of the ionized regions at the epoch when the cross-correlation changes sign.}
    \label{BSD}
\end{figure}

To validate the end-of-reionization estimates from the 21cm--[CII] cross-correlation, we examine the evolution of the mean free path (MFP) of ionizing photons, which traces the growth and connectivity of ionized regions. This validation test assesses whether the MFP of ionizing photons at the end of reionization becomes comparable to the size of the simulation box. We investigate the evolution of the MFP of ionizing photons during the late stages of reionization, using the publicly available \texttt{tools21cm}\footnote{\url{https://tools21cm.readthedocs.io/installation.html}} package \citep{Giri_2020} to estimate the bubble size distribution (BSD) from our simulated ionization fields. We calculate the BSD at seven redshifts spanning $5.7\leq z\leq7.0$ and show the results for $z\simeq7.0$, $6.5$, and $5.7$ in figure \ref{BSD}(Left). We estimate the MFP from the radius $R$ at which the BSD reaches its maximum, with the vertical lines in the figure indicating the corresponding peak positions\footnote{In this work, we use the term MFP to denote the characteristic size of ionized bubbles, as estimated from the peak of the bubble size distribution. This definition does not account for the additional attenuation of ionizing radiation due to the optical depth of high-density regions within the ionized bubbles}. We evaluate the MFP at the redshifts associated with the transitions from negative to positive 21cm--[CII] cross-power. 

Figure \ref{BSD}(Left) shows a substantial increase in the characteristic MFP as redshift decreases from $z\simeq7.0$ to $z\simeq5.7$. This trend is consistent with the growth and merger of ionized regions as reionization progresses. As these regions expand and begin to percolate, ionizing photons can propagate over increasingly large distances before encountering neutral regions. The increasing MFP therefore provides a complementary indication of the evolving topology of the ionization field toward the end of reionization.

We further examine the relationship between the MFP and the transition redshift, $z_{\rm cross}$, identified from the 21cm--[CII] cross-power spectrum at different Fourier modes. As shown in figure \ref{BSD}(Right), the transition occurs first at large $k$, corresponding to small spatial scales, and at progressively lower redshifts toward smaller $k$, which probes larger spatial scales. This behavior is consistent with the fact that the evolution of the cross-correlation on large scales reflects the development of extended ionized regions and the approach toward global reionization. As a result, the MFP associated with the transition increases toward lower transition redshifts, approaching the size of the simulation volume. This should be kept in mind when interpreting the inferred MFP, as finite-volume effects may become important when the characteristic scale approaches the simulation volume.

    
\subsection{Detectability estimates} 
\label{Detectability Estimates}
To study the observational prospects of detecting the predicted cross-correlation for upcoming LIM surveys. For the SKA-Low AA$^{*}$ mock observations, we adopt a total observing time of $1000$ hours, with an integration time of $60$ seconds. The observations are assumed to consist of $8$-hour tracks centered on the Meridian crossing of the target field. We consider a coeval simulation volume of $215\,\mathrm{Mpc}$ on a side, discretized on a $384^3$ grid, yielding a spatial resolution of $\Delta x \simeq 0.56\,\mathrm{Mpc}$ per voxel. At each redshift, the central observing frequency is determined from $\nu=1420/(1+z)\,\mathrm{MHz}$. The frequency resolution associated with a single line-of-sight voxel and the total frequency bandwidth corresponding to the $215\,\mathrm{Mpc}$ depth of the simulation volume are then calculated from the cosmological redshift--distance relation, which is used to calculate $k_{\parallel}$. The corresponding angular resolution and angular extent of the simulated volume are calculated in the same manner, and the latter is compared with the SKA-Low Field of View (FoV) to ensure that the simulated patch is fully contained within the primary beam. The detailed prescription for these quantities is described in section \ref{Detectability Estimation}.

\begin{figure}
    \centering
    \includegraphics[width=1\linewidth]{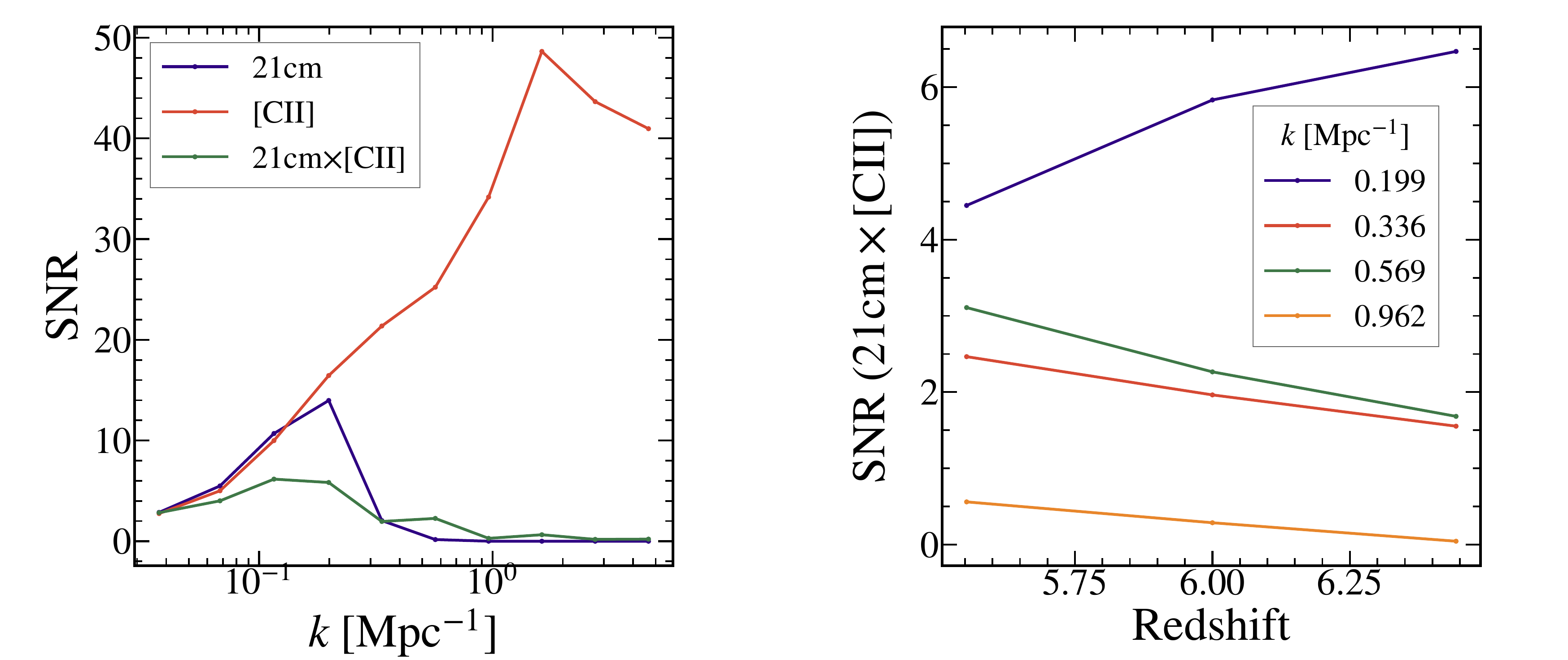}
      \caption{\textbf{Left:} Signal-to-noise ratio (SNR) as a function of wavenumber $k$ at $z\sim6$. The blue, red, and green curves show the SNR of the 21cm auto-power spectrum, the [CII] auto-power spectrum, and the 21cm--[CII] cross-power spectrum, respectively. \textbf{Right:} 21cm--[CII] SNR as a function of redshift for different $k$-modes in the range $\rm 0.2$--$\rm 0.96~\rm{Mpc}^{-1}$.}
    \label{SNR}
\end{figure}

For the FYST Deep Sky Survey (DSS), we use the survey specifications adopted in \cite{CCAT_2023}. The survey has a total observing time of $4000$ hours and covers an area of  $4\,\mathrm{deg}^2$. The corresponding instrumental thermal noise is then calculated following the prescription described in section \ref{Detectability Estimation}. Figure \ref{SNR}(Left) shows the SNR of the 21cm auto-power spectrum, the [CII] auto-power spectrum, and the 21cm--[CII] cross-power spectrum as a function of $k$-modes, at $z\sim6$.

At $z\sim6$, the 21cm auto-power spectrum SNR is limited by sample variance at large scales (small $k$) and by instrumental noise at small scales (large $k$), peaking at $k\simeq0.2\,{\rm Mpc}^{-1}$ with ${\rm SNR}\simeq14$, reaching $\simeq3$, $\simeq6$, and $\simeq11$ at $k\simeq0.04$, $0.07$, and $0.15\,{\rm Mpc}^{-1}$, respectively, before falling below unity beyond $k\gtrsim0.5\,{\rm Mpc}^{-1}$. The [CII] auto-power spectrum SNR again rises steadily with $k$, reaching $\simeq48$ near $k\simeq1.7\,{\rm Mpc}^{-1}$; because the white noise is scale-independent, the SNR varies only due to the scale dependence of signal power and sample variance. The 21cm--[CII] cross-power spectrum peaks at ${\rm SNR}\simeq6.2$ near $k\simeq0.1\,{\rm Mpc}^{-1}$ and remains detectable out to scales where the 21cm auto-power spectrum SNR has already dropped to zero. This behavior highlights a key advantage of the 21cm--[CII] cross-power spectrum: across the range of scales relevant to this work, the cross-power spectrum SNR remains significant even at scales where the 21cm auto-power spectrum SNR has already dropped below unity.

Figure~\ref{SNR}(Right) shows the redshift evolution of the SNR for our targeted $k$-modes. At $z\sim5.5$, the noise variance is lower than at $z\sim6$ and $z\sim6.5$, which leads to a higher SNR. For $k\simeq0.199\,\mathrm{Mpc}^{-1}$, however, the signal power at $z\sim6.5$ is higher than at $z\sim5.5$, and the SNR is correspondingly higher at $z\sim6.5$. The variation of the SNR with redshift across different $k$-modes thus arises from the interplay between the evolution of the signal power and that of the noise variance.
Over this redshift range, the cross-power spectrum is thus detectable with high significance over the full range of $k$-modes relevant to identifying the 21cm--[CII] correlation signal, demonstrating that this observable should be robustly detectable with upcoming SKA-Low and FYST observations.

    
For more details on detectability forecasts with SKA-Low and FYST, interested readers may refer to \citep{Abinash_2019, Samit_2025} and \citep{Anirban_2025}, respectively.

\section{Summary and conclusion}
\label{Summary}
In this work, we investigate for the first time the potential of the cross-correlation between the redshifted HI 21cm signal and [CII] line emission to probe the end of the EoR. We explore whether the evolving spatial relationship between the 21cm signal tracing HI and [CII]-emitting galaxies encodes a distinctive signature of the end of reionization, and evaluate the prospects for detecting this signature with upcoming LIM observations. For this purpose, we employ a modified version of the \texttt{ReioniYuga} simulation that incorporates a halo-based HI prescription allowing us to generate 21cm signal from both the IGM and halo, and also generate [CII] LIM fields. We characterize their evolving relationship using PCC and cross-power spectrum, and further investigate the detectability of the cross-correlation for SKA-Low and FYST observations.

We first examine the evolution of the cross-correlation using the PCC of 21cm brightness temperature and [CII] LIM fields, shown in figure \ref{cross_variance}. The PCC successfully captures the overall evolution of the correlation and provides an estimate of the redshift at which reionization is completed. However, we find that the inferred transition depends on the smoothing scale adopted in the analysis, indicating that the PCC does not contain any explicit scale information. To overcome this limitation, we investigated the scale-dependent 21cm--[CII] cross-power spectrum, shown in figure \ref{cross}. We define the completion of reionization as the redshift at which the cross-power spectrum becomes positive across all scales considered in our analysis for the first time. This criterion is physically motivated by the fact that, once the diffuse IGM is almost completely ionized, both the 21cm and [CII] signals predominantly trace the underlying galaxy distribution, resulting in a positive correlation over the full range of observable scales.

To validate this interpretation, we independently examined the characteristic size of ionized regions through the bubble size distribution and estimated the corresponding photon mean free path. We find a substantial increase in the mean free path toward the inferred end-of-reionization redshift, reaching $\lambda_{\rm mfp} \simeq 151~{\rm Mpc}$ at $z \simeq 5.7$. This indicates the emergence of large, interconnected ionized structures and is consistent with the completion of reionization inferred from the cross-correlation analysis. 

Determination of the end of reionization redshift is subjective. It may give different answers when one follows different criteria than those used in this work. A global neutral fraction threshold may yield different results compared to the change in sign of the cross-correlation coefficient from negative to positive at different $k$-modes. Our definition of the end of reionization is such that, when for the first time the cross-power spectrum of 21cm and [CII] becomes positive at the smallest $k$-mode (i.e. the largest observed scale) considered and also remains positive at all other observed $k$-modes, i.e., $0.2 \leq\,k\leq 1\, {\rm Mpc^{-1}}$. Following this definition, our analysis yields $z_{\rm end} \sim 5.748$ at which $\rm \bar x_{HI}^{M}=0.008$. The crossover of the 21cm--[CII] arises from the competition between an anti-correlated diffuse-IGM component and a positively correlated halo-associated HI component, with the scale dependence of their crossover encoding information about the evolving morphology of the ionized IGM.

Finally, we investigated the detectability of the proposed signal using survey specifications representative of SKA-Low AA$^{*}$ array configuration \citep{sridhar_2025} and FYST \citep{CCAT_2023}. By incorporating realistic instrumental noise realizations of these experiments, we estimated the SNR for the 21cm auto-power spectrum, the [CII] auto-power spectrum, and the 21cm--[CII] cross-power spectrum, shown in figure \ref{SNR} at redshift $z\sim6$. We find that the relevant $k$-modes required to identify the sign transition are detectable with a significant signal-to-noise ratio, demonstrating that future intensity-mapping surveys will be capable of using the methodology proposed in this work to estimate the end-of-reionization redshift. 

We emphasize that the results presented here are specific to one fiducial reionization and [CII] emission model, and the inferred end-of-reionization redshift should be interpreted within the framework of those models. Moreover, our analysis does not include line interlopers, astrophysical foregrounds, or instrumental systematics beyond thermal noise and cosmic variance. In future work, we will incorporate these effects and assess whether the proposed method for determining the end-of-reionization redshift remains robust under realistic observational conditions. We also plan to repeat the analysis for a range of reionization and [CII] LIM models, in order to quantify how the inferred end-of-reionization redshift varies with the underlying assumptions about the sources, the reionization history, and the [CII] luminosity prescription, and thereby to test the general applicability of the proposed criterion.



\acknowledgments
SA thanks Vishrut Pandya for useful discussions. LN acknowledges financial support from the Department of Science and Technology, Government of India, through the INSPIRE Fellowship [IF210392] and from the Abdus Salam International Centre for Theoretical Physics (ICTP) under the ‘ICTP Sandwich Training Educational Programme (STEP)’ SMR.3991,
SMR.4129 and SMR.4263. YM acknowledges financial support from the Department of Science and Technology, Government of India, through the INSPIRE Fellowship [IF220304]. CSM acknowledges financial support from the ARCO Prize Fellowship. We acknowledge the use of computing infrastructure for this work, which is hosted at the DAASE, IIT Indore, and was procured through funding via the Department of Science and Technology, Government of India sponsored DST-FIST grant No. SR/FST/PSII/2021/162 (C) awarded to the DAASE, IIT Indore.

\appendix
\section{Appendix}
\label{Appendix}
\label{Supplementary}

The IGM-only 21cm--[CII] cross-power spectrum is negative at almost all scales, except for a small positive feature at $k\simeq0.33\,\mathrm{Mpc^{-1}}$. We tested whether this feature arises from binning of the $k$-modes, or due to a specific statistical realization of the simulated Universe, or due to the outlier $k$-modes.

\begin{figure}
    \centering
    \includegraphics[width=1\linewidth]{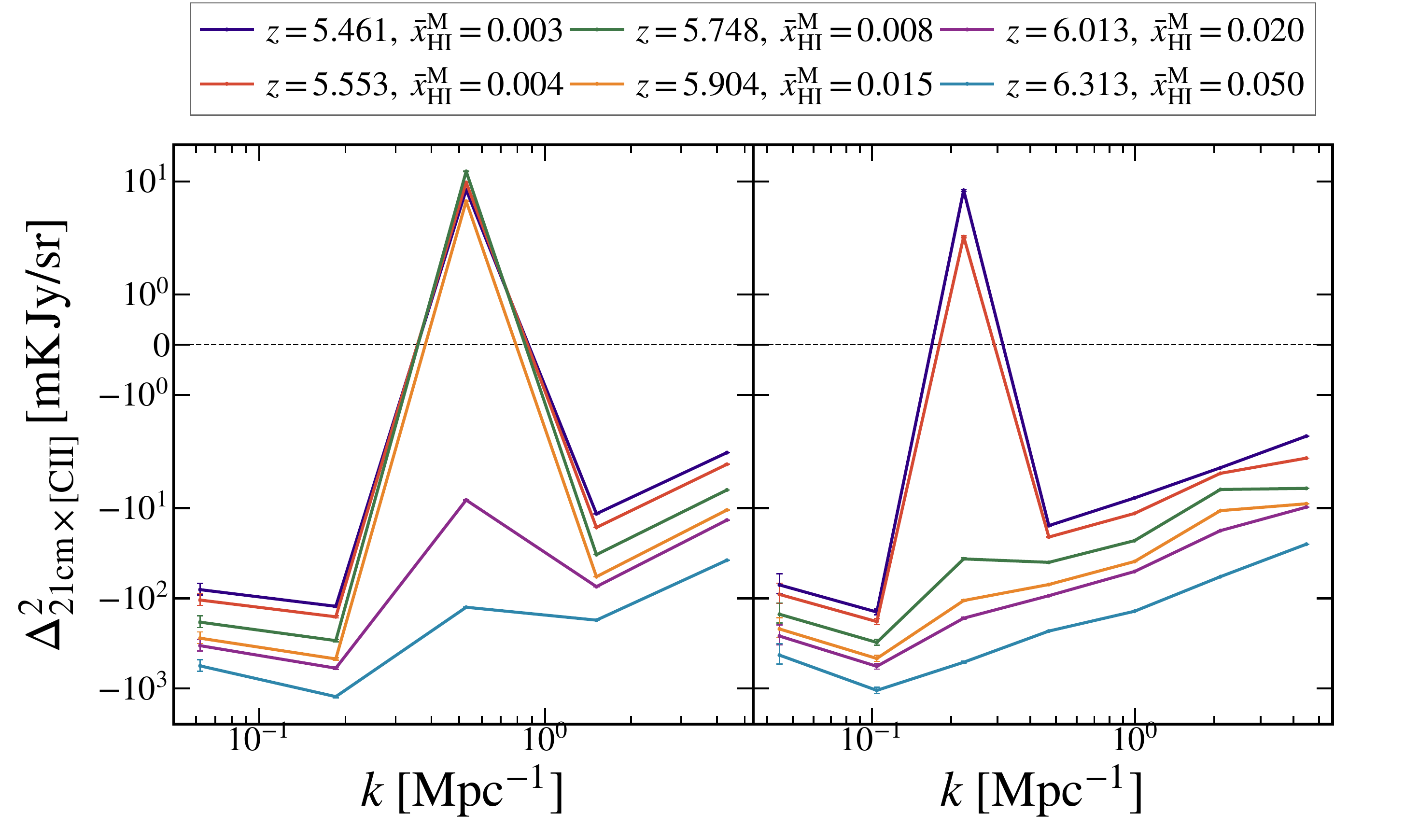}
      \caption{\textbf{Left:} IGM only 21cm--[CII] cross-power spectrum for $5$ $k$-bins. \textbf{Right:} Same as (Left) but for $7$ $k$-bins.}
    \label{IGM_CII_bin}
\end{figure}

\begin{figure}
    \centering
    \includegraphics[width=1\linewidth]{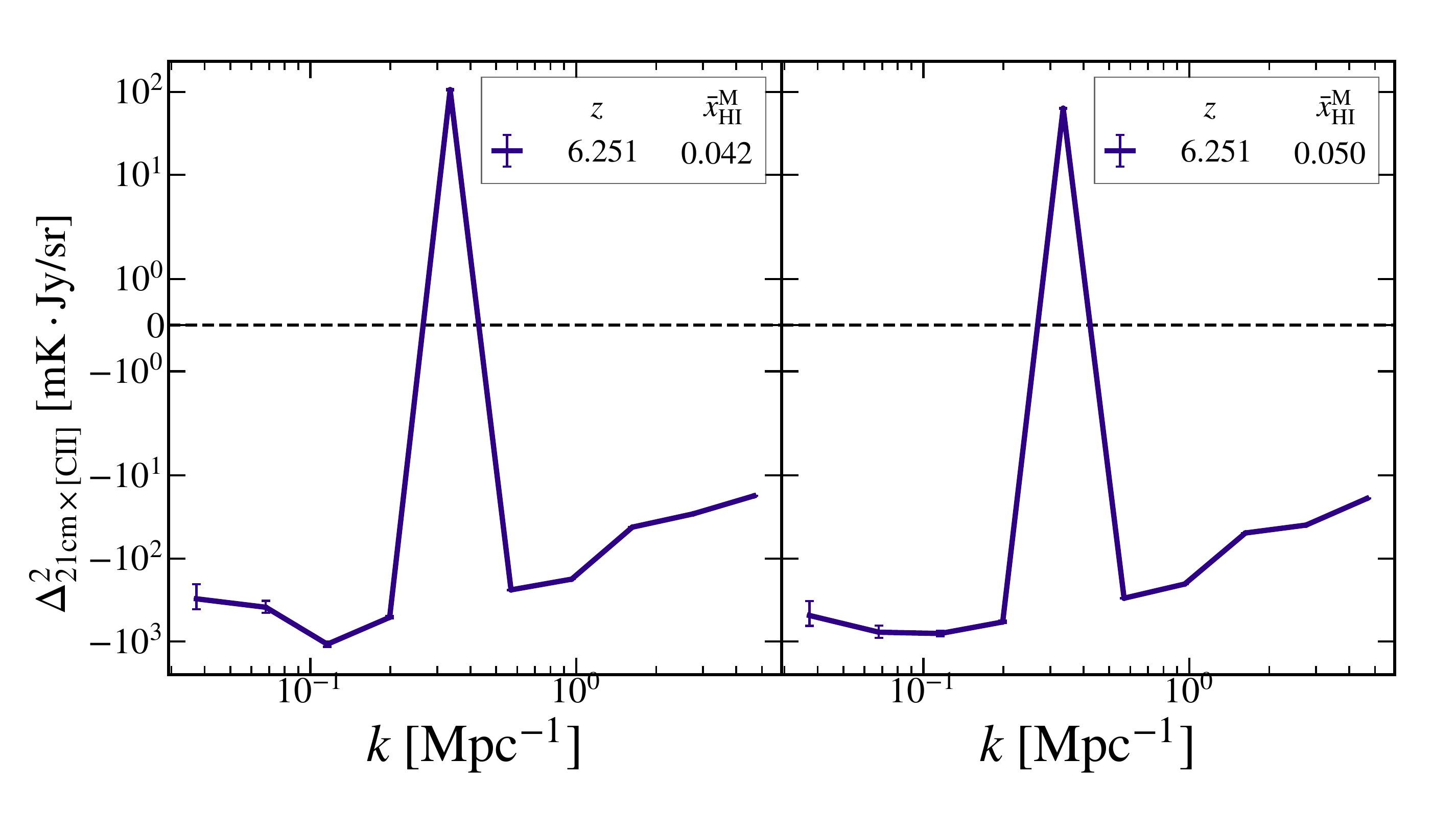}
      \caption{\textbf{Left:} IGM-only 21cm--[CII] cross-power spectrum. \textbf{Right:} Same as (Left) plot but for a statistically different realization of the Universe.}
    \label{IGM_CII_seed}
\end{figure}

First, we varied the binning. Figure \ref{IGM_CII_bin} shows the cross-power spectrum computed with different numbers of logarithmically spaced $k$-bins [$N_{\rm bin}=5,\,{\rm and}\, 7$]. We observe that the same feature is present at around the same wavenumber in both cases; thus, it does not depend on the choice of bin size. Second, we repeated the estimation of cross-power spectra for two statistically independent realizations of our Universe while keeping the cosmological and reionization parameters identical (figure \ref{IGM_CII_seed}). The feature appears in both realizations at the same wavenumber, indicating that it is not a fluctuation tied to a particular realization of the density field.

\begin{figure}
    \centering
    \includegraphics[width=1\linewidth]{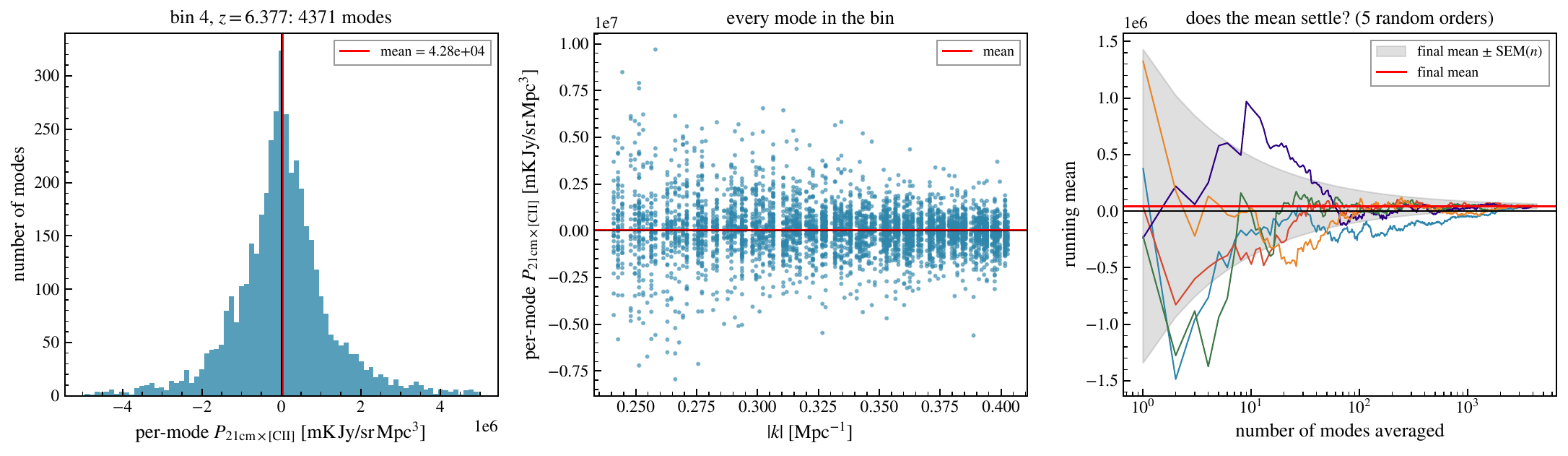}
    \caption{\textbf{Left:} Distribution of the cross-power spectrum values of all the modes present in the $k$-bin centered at $0.336\, \rm Mpc^{-1}$. \textbf{Middle:} Distribution of the cross-power spectrum values of all the modes over the $k$-mode range centered at $0.336\, \rm Mpc^{-1}$. \textbf{Right:} For $5$ random orderings of the average of all the modes in the $k$-bin centered at $0.336\, \rm Mpc^{-1}$.}
    \label{Bin_4_distribution}
\end{figure}

Third, we examined the individual Fourier modes contributing to the $k$-bin that contains the feature. For each of the $N_{\rm modes}=4371$ independent modes in this bin, we computed their contribution to the cross-power spectrum estimation, $\mathrm{Re}\,[\delta{T}_{b}^{*}(\bm{k})\,\tilde{I}_{\rm CII}(\bm{k})]$, whose average over the bin gives the binned estimate of the cross-power spectrum. The distribution of these contributions (left panel of figure \ref{Bin_4_distribution}) is broad and approximately symmetric around their mean, and no small set of outlying modes dominates the bin average estimate. The mean is positive. We also verified that the estimate is stable with respect to the order in which the modes are averaged: the running mean converges to the bin average for random orderings of the modes (right panel of figure \ref{Bin_4_distribution}).
    

These tests show that the feature is not an artifact of the binning, of the particular realization, or of a few outlying modes. Its physical origin, however, remains unclear and is left for future work.

\bibliographystyle{unsrturl}
\bibliography{biblio}
\end{document}